**Field-Tailoring Quantum Materials via Magneto-Synthesis:**

**Metastable Metallic and Magnetically Suppressed Phases in a Trimer Iridate**

Tristan R. Cao[1], Hengdi Zhao[1], Xudong Huai[2], Arabella Quane[1], Thao T. Tran[2], Feng Ye[3] and Gang Cao[1]

[1]Department of Physics, University of Colorado at Boulder, Boulder, CO 80309

[2]Department of Chemistry, Clemson University, Clemson, SC 29634

[3]Oak Ridge National Laboratory, Oak Ridge, TN 37830

We demonstrate that applying modest magnetic fields (< 0.1 T) during high-temperature crystal growth can profoundly alter the structure and ground state of a spin-orbit-coupled, antiferromagnetic trimer lattice. Using $BaIrO_3$ as a model system, whose ground state is intricately dictated by the trimer lattice, we show that *magneto-synthesis*, a field-assisted synthesis approach, stabilizes a structurally compressed, metastable metallic and magnetically suppressed phases inaccessible via conventional methods. These effects include a 0.85% reduction in unit cell, 4-order-of-magnitude decrease in resistivity, a 10-fold enhancement of the Sommerfeld coefficient, and the collapse of long-range magnetic order -- all intrinsic and bulk in origin. First-principles calculations confirm that the field-stabilized structure lies substantially above the ground state in energy, highlighting its metastable character. These large, coherent and correlated changes across multiple bulk properties, unlike those caused by dilute impurities, defects or off-stoichiometry, point to an intrinsic field-induced mechanism. The findings establish magneto-synthesis as a powerful new pathway for accessing non-equilibrium quantum phases in strongly correlated materials.

## I. Introduction

Materials synthesis lies at the heart of discovery and control of novel quantum and functional materials that underpin future technologies. The grand materials challenge is the rational design and synthesis of new solids with targeted structures and functionalities. However, achieving this ambition has proven difficult, not for lack of effort, but mainly due to the lack of effective synthesis tools to finely control the thermodynamic energy landscape that intricately governs the phase formations and detailed local crystal distortions that dictate the performance of quantum materials. Historically, synthesis has relied on thermal and mechanical control, yet many predicted quantum phases, such as quantum spin liquids, topological superconductors, or high-temperature superconductors, remain unrealized, suggesting the need for new synthesis paradigms [e.g., 1-10].

Among the few available approaches to alter energy landscapes during synthesis, high-pressure techniques have been widely used to stabilize metastable phases or induce novel electronic states in quantum materials [11, 12]. However, high-pressure synthesis typically requires pressures of several giga pascal (1GPa ~ $10^4$ atmospheres) or higher, thus small-volume apparatus and often yields polycrystalline materials, limiting its broader applicability. Moreover, high-pressure materials suffer from the destabilization after pressure release [12].

One emerging strategy is to introduce magnetic fields as active tuning parameters during synthesis, in what we term **magneto-synthesis (Figs.1a-1b)** [13]. Field-assisted synthesis offers a scalable, directional, and non-contact method for tuning thermodynamic and kinetic pathways during crystal growth, in sharp contrast to high-pressure synthesis. In particular, magnetic fields can fundamentally reshape the thermodynamic and kinetic energy landscape of solid-state reactions [13-28], offering multiple advantages: suppression of convection and temperature gradients, stabilization of metastable or highly ordered structures, and the modification of free energy landscapes to favor otherwise inaccessible phases (see **Table 1**).

| **Table 1. Key advantages of field-assisted synthesis and processing** |
|---|
| *1. New phases and structures* |
| *2. Highly ordered (micro)structures* |
| *3. Acceleration of phase formation* |
| *4. Stabilization of non-equilibrium metastable phases* |
| *5. Reduction of lattice distortions* |
| *6. Spin micromechanics* |
| *7. Large magnetic domains* |
| *8. Low convection, temperature fluctuations and striations* |
| *9. High homogeneity* |

While early applications of magneto-synthesis focused primarily on semiconductors and steels [19–25, 28], its use in correlated electron systems has only recently gained traction. Our recent studies [13, 26, 27] have helped extend this method into the quantum materials domain, demonstrating that magneto-synthesis is particularly effective in modulating ground states of materials with strong spin-orbit interactions (SOI) and pronounced magnetoelastic coupling. Systems such as the spin-orbit-assisted Mott insulator $Sr_2IrO_4$, the spin liquid $Ba_4Ir_3O_{10}$, and the lattice-driven Mott insulator $Ca_2RuO_4$ exhibit profound changes in structural and physical properties when synthesized under weak magnetic fields [13, 26, 27].

Prior work, including our own, has identified three key mechanisms that underpin magneto-synthesis of spin-orbit-coupled materials [13, 25, 26] : **(1)** Lorentz-force suppression of melt convection and inhomogeneity, **(2)** magnetic-field tuning of phase stability via $\mathbf{\Delta G_{Tot} = \Delta G_{Chem} + \Delta G_B}$, where $\Delta G_{Chem}$ is Gibbs free energy in the absence of applied magnetic field B, and $\Delta G_B$ due to B, and **(3)** amplification of magnetoelastic and SOI effects that couple magnetic and structural order. At first glance, such effects might seem negligible during high-temperature synthesis, as the magnetic energy (~ $10^{-5}$ eV for $\mu_0 H = 0.1$ T) is vastly smaller than the thermal energy (> 0.1 eV at 1000 °C). However, this intuition overlooks the fact that **crystal growth is governed not only by thermal averages but also nonequilibrium kinetics,** interfacial energies, and local structural fluctuations. Even weak magnetic fields can subtly reshape free-energy landscapes, bias nucleation, and influence melt dynamics. As demonstrated in our earlier work [13], these small perturbations can steer synthesis outcomes in materials with strong spin-orbit and magnetoelastic interactions by suppressing convection, amplifying lattice-spin coupling, and stabilizing otherwise inaccessible metastable phases. Despite the low absolute energy scale, magnetic fields can thus exert a **disproportionate influence** on phase formation in quantum materials near structural or magnetic instabilities.

Here, we show that these principles can be systematically harnessed in a model system, $BaIrO_3$, a prototypical spin-orbit-coupled trimer-lattice antiferromagnet [29-51]. Trimer lattices consist of three tightly bonded transition-metal ions (**Figs. 1c-1d**), forming molecular-like units that support internal electronic and structural degrees of freedom beyond those found in conventional crystal-field environments [32, 33]. These internal modes, especially the average **metal-metal bond distance** within trimers (**Fig. 1c**), serve as highly sensitive **tuning knobs** for modulating physical behavior in high-Z (Z being atomic number) compounds such as iridates and ruthenates.

As summarized in **Fig. 1e**, an empirical phase diagram compiled from our broader studies of trimer lattices [13, 26, 34–40, 43, 45–50] reveals a clear correlation between the magnetic ordering temperature $T_N$ and the average metal-metal bond distance. Magnetic order collapses once this distance contracts below ~ 2.58 Å, a regime where enhanced direct overlap between transition-metal orbitals destabilizes conventional exchange pathways and drives magnetic frustration [33, 34]. However, conventional synthesis methods have proven ineffective in precisely tuning trimer lattice parameters, motivating our use of magneto-synthesis as a means to structurally tailor these high-Z quantum materials.

These structural and electronic features, combined with the sensitivity of trimer lattices to field-induced perturbations, make $BaIrO_3$ an ideal platform to investigate how weak magnetic fields can steer synthesis toward **metastable phases** and tailor material properties **beyond equilibrium thermodynamic limits**.

At ambient conditions, $BaIrO_3$ exhibits an unusual coexistence of charge density wave (CDW) order and antiferromagnetism with a Néel temperature of 185 K and a sizable insulating gap of ~ 0.1 eV [43, 44]. Its complex magnetic structure arises from trimer-based spin canting and strong Dzyaloshinskii-Moriya interactions. Further studies have uncovered evidence for orbital magnetism [48], making $BaIrO_3$ a promising candidate for unconventional magnetic and electronic phases. Using only permanent magnets producing fields < 0.1 T inside the furnace chamber, we observe dramatic structural contraction by up to 0.85%, suppressed magnetism by more than 100 K, and a transition to an emergent metallic state - **an effect analogous to or even stronger than applying several GPa of hydrostatic pressure** [50, 52]. These results validate magneto-synthesis as a powerful new tool to access metastable quantum phases, motivating the design and development of advanced field-assisted growth platforms, such as our upcoming laser-heated 5 T system [53].

It is worth noting that the magnitude and consistency of the observed changes across structure, resistivity, magnetism, and heat capacity together with first-principles calculations rule out impurity effects and establish magneto-synthesis as the intrinsic driver.

In addition, to further support the intrinsic nature of the field-induced transformations observed in $BaIrO_3$, we present **Supplementary Figs. 1-10** for additional x-ray diffraction and magnetic susceptibility data collected from independently grown crystals corresponding to the M1, M2, and M2* synthesis conditions (see definitions of M1, M2 and M2* below). These measurements were performed on multiple batches synthesized under identical magnetic field protocols. The results confirm the high reproducibility of the magnetic behavior reported in the main text, including the systematic suppression of the Néel temperature $T_N$ with increasing synthesis field strength. The close agreement across samples further rules out sample-specific anomalies or stochastic impurity effects, reinforcing the conclusion that the observed phenomena are robust consequences of magneto-synthesis.

**II. Experimental**

The field-tailored single crystals of $BaIrO_3$ were grown in a box furnace capable of reaching temperatures up to 1500 °C. For the control studies, growth was conducted in the presence of either one permanent magnet (denoted as M1) or two magnets (M2), each rated at 1.4 T and securely attached to the exterior of the furnace (**Fig. 1a**). Because magnetic field strength decays with distance d according to a $1/d^3$ relationship, the actual field inside the furnace chamber was significantly lower. Using a Gaussmeter, we measured the internal field at the crucible position to be in the range of 0.01-0.04 T for M1 and 0.02-0.06 T for M2 (**Fig.1b**). The field strength varied depending on the exact placement of the crucible; samples denoted as M2* were grown in a different vertical position within the furnace than those labeled M2. Non-tailored single crystals were synthesized under otherwise identical thermal conditions but without the application of any magnetic field. **All results reported here were reproduced across multiple independently synthesized batches, ensuring consistency and reliability (Supplementary Figs. 6-10)**. The crystals obtained from both field-tailored and non-tailored conditions were confirmed to be of high structural quality and chemical purity through single-crystal X-ray diffraction (SCXRD), energy-dispersive X-ray spectroscopy (EDX), and scanning electron microscopy (SEM) (**Supplementary Information Section III, and Tables S1 and S2**). Magnetic, electrical transport, and thermal properties were characterized using a Quantum Design 14 T Dynacool Physical Property Measurement System (PPMS) and a 7 T Magnetic Property Measurement System (MPMS-7 SQUID magnetometer). Additional details on synthesis, measurement procedures and computational details are provided in the **Methods** section.

**III. Results and discussion**

$BaIrO_3$ crystallizes in the monoclinic space group *C2/m* and features a trimer lattice composed of three face-sharing $IrO_6$ octahedra that form $Ir_3O_{12}$ units [41-44] (**Fig. 1c**). These trimers are further linked via corner-sharing $IrO_6$ octahedra, forming quasi-one-dimensional chains along the crystallographic *c*-axis (**Fig. 1d**). A characteristic monoclinic distortion leads to twisting and buckling of the trimer chains, quantified by the Ir-O-Ir bond angle, θ, between neighboring corner-sharing octahedra (**Fig.1d**). This distortion plays a critical role in stabilizing the insulating antiferromagnetic (AFM) ground state of $BaIrO_3$ [43, 46, 47]. At 100 K, the bond angle θ is measured to be 155.5° in the non-tailored crystal but increases to 157.7° in the field-tailored M2 sample (**Supplementary Fig.1**), closer to the undistorted bond angle 180°, an effect with significant implications for the physical properties, as discussed below.

Magneto-synthesis also leads to a notable contraction of the Ir-Ir bond length within the trimer, specifically the Ir1–Ir4 distance, $d_{Ir1-Ir4}$, by as much as 0.69% (**Fig. 2a**). This bond shortening strengthens the interactions among the three Ir ions and directly influences the magnetic ground state, as discussed above **(Fig.1e**). Detailed analysis of the lattice parameters across all field-tailored samples reveals a consistent shrinkage of the *a*- and *c*-axes, accompanied by a slight elongation of the *b*-axis (see **Supplementary Fig. 2**). Collectively, these structural modifications yield a remarkable reduction in unit cell volume *V* by up to 0.85% (**Fig. 2b**).

Remarkably, the structural compression, suppression of long-range order, and a transition to an emergent metallic state achieved here via magneto-synthesis (< 0.1 T) mimic or *exceed* those previously reported under **multi-GPa hydrostatic pressures** [50, 52]. This indicates that magneto-synthesis effectively stabilizes a high-energy, metastable phase otherwise inaccessible through conventional synthesis. These findings are summarized in **Table 2**. **Such enormous bulk effects also conclusively rule out any extrinsic cause**.

Under ambient conditions, the magnetic structure of $BaIrO_3$ is governed by SOI and crystal distortions. Ir spins preferentially align along the *c*-axis, the magnetic easy axis (**Fig. 2c**). Within each trimer, the two outer Ir spins align parallel to each other, while the central Ir spin aligns antiparallel, forming a net canted spin structure. This canting, amplified by the trimer distortion and Dzyaloshinskii–Moriya interactions, gives rise to weak ferromagnetism superimposed on the AFM background [41-44].

The *c*-axis magnetic susceptibility $\chi_c$ reveals a progressive suppression of the Néel temperature $T_N$ with increasing magnetic field strength during synthesis (**Fig. 2d**). Specifically, $T_N$ decreases from 185 K in the non-tailored crystal to 150 K (M1), 115 K (M2), and becomes indistinct in the M2* sample. These observations are consistent with the empirical phase diagram **Fig. 1e**. As $d_{Ir1\text{-}Ir4}$ bond distance in the field-tailored $BaIrO_3$ approaches the threshold ~2.58 Å, magnetic order is nearly extinguished underscoring the pivotal role of trimer bond contraction in destabilizing long-range magnetic order.

The weak ferromagnetism present in the non-tailored crystal manifests as pronounced magnetic hysteresis and domain behavior, as seen in the isothermal magnetization data where $M_a$ >

$M_c$ (**Fig. 2e**). In contrast, the field-tailored M2 sample exhibits negligible hysteresis (**Fig. 2f**), indicating reduced spin canting and enhanced domain mobility. This reduction in magnetic anisotropy and hysteresis directly reflects the structural simplification and enhanced symmetry of the field-stabilized phase.

Consistent with the structural and magnetic changes discussed above, the electrical resistivity $\rho_c$ for the *c* axis (on a logarithmic scale) exhibits a dramatic decrease by up to 4 orders of magnitude in the field-tailored M2 sample, indicating a clear insulator-to-metal transition (**Fig. 3a**). This emergent metallic state is nearly unaffected by applied magnetic fields, in sharp contrast to the insulating behavior of the non-tailored sample, which shows a large positive magnetoresistance due to field-enhanced localization (**Fig. 3b**). Notably, this transition is not induced by external fields during measurement but results from magneto-synthesis itself, which stabilizes a metastable, compressed phase with reduced structural distortion (**Figs. 2a-2b**). The shortened Ir–Ir bond distance within the trimer enhances inter-site electron interactions, and the significant contraction of the unit cell volume (up to 0.85%) naturally increases the electronic bandwidth and density of states at the Fermi level, both of which favor metallicity.

Further support for this emergent phase comes from bulk thermodynamic measurements. The low-temperature specific heat, C(T), reveals marked differences between the non-tailored and field-tailored samples, consistent with the onset of metallic behavior. The total specific heat can be expressed as: $C(T) = \gamma T + \beta T^3 + \beta' T^\alpha$. The first term $\gamma T$ accounts for the electron contribution to C, where $\gamma$ is the Sommerfeld coefficient related to the density of states near the Fermi surface. The second term $\beta T^3$ arises from the phonon contribution, where $\beta$ represents the inverse of the thermal energy. The third term $\beta' T^\alpha$ pertains to the magnetic contribution, where $\beta'$ is associated with magnons and $\alpha$ is typically 3 and 3/2 for three-dimensional antiferromagnets and ferromagnets, respectively.

The pronounced peak in C(T) at $T_N$ = 150 K for the field-tailored M1sample manifests a strongly ordered state (**Fig.3c**), leading to a significant entropy removal $\Delta$S (**Fig.3c,** right scale). The prominence of this magnetic peak at such a high temperature is noteworthy, as the phonon contribution to C(T) scales with $T^3$ and is expected to be dominant at high temperatures, such as 150 K. The presence of the large magnetic peak implies that the AFM state, which also scales with $T^3$, is highly ordered, making an unusually large contribution to C(T) in this temperature range, particularly in comparison to the non-tailored sample where the magnetic anomaly related to $T_N$ is hardly discernible (**Fig.3c**). The contrasting behaviors provide further evidence of a significantly higher degree of order in the field-tailored M1 sample.

To extract $\gamma$, we plot C/T versus $T^2$ below 10 K (**Fig. 3d**), yielding $C/T = \gamma + (\beta+\beta')T^2$. Both the non-tailored and field-tailored M1 samples display linear behavior in this regime, indicating that the heat capacity is dominated by phonons and magnons. However, the slope $(\beta + \beta')$ differs significantly: 0.055 mJ/mol·$K^4$ for the non-tailored sample and 0.127 mJ/mol·$K^4$ for M1, reflecting changes in both lattice and magnetic contributions due to structural modification. Crucially, the extrapolated intercepts at $T^2 = 0$ yield Sommerfeld coefficients of $\gamma = 1.3$ and 12.0 mJ/mol·$K^2$, respectively. The negligible $\gamma$ in the non-tailored sample confirms its nonmetallic nature, while the significant $\gamma$ for M1 is consistent with metallic behavior and an enhanced density of states at the Fermi level. Even more striking is the behavior of the M2* sample, which deviates strongly from linearity, consistent with the complete suppression of long-range AFM order (**Fig. 2d**) and the stabilization of a highly correlated metallic state. The extrapolated $\gamma$ value reaches 35.0 mJ/mol·$K^2$, indicating a substantially increased density of states and electronic correlations, a hallmark of a strongly correlated metallic ground state [54]. It is noteworthy that heat capacity is a bulk effect and insensitive to small impurities or defects. The substantial changes in C further confirm the intrinsic effect of magneto-synthesis.

To further understand the impact of field-tailored synthesis on the structural and electronic stability of $BaIrO_3$, we performed a systematic comparison of five structural models using density functional theory (DFT) calculations: the fully relaxed ground state, a position-relaxed structure under constrained lattice parameters, a non-tailored experimental structure, and two field-tailored structures (M1 and M2) obtained under synthesis in external field conditions. The calculated density of states (**Supplementary Fig. 3**) shows an increased number of states at the Fermi level ($E_F$), indicating that $BaIrO_3$ becomes more metallic through field-tailored synthesis. Additionally, the broader energy distribution in the field-tailored structures suggests improved orbital overlap. The evolution of electronic properties is summarized in **Fig. 4**.

Structurally, **Fig. 4a** shows a monotonic decrease in unit cell volume from the fully relaxed to the field-tailored configurations, accompanied by changes in the lattice parameters *a*, *b*, and *c* axis (right scale). This compaction leads to a significant increase in internal force and stress (**Fig. 4b**), indicating that the system is increasingly strained under non-equilibrium conditions. Along with the compression, charge redistribution takes place. **Fig. 4c** presents the Mulliken charges on the Ir-5*d* and Ba-6*s* orbitals, showing a shift in electron density from Ir to Ba with increasing constraint, confirming the redistribution that affects the electrostatic environment and bonding interactions. To assess the electronic instability, we calculated the density of energy (DOE). The negative DOE values near the Fermi level suggest a destabilizing character (**Supplementary Figs. 4-5**). This is further illustrated by the integrated DOE in **Fig. 4d** (right scale), which shifts toward greater band energy under constraint, with the field-tailored structures exhibiting the highest electronic instability. Finally, **Fig. 4d** also shows the total energy differences relative to the fully relaxed structure. **The field-tailored configuration lies ~ 18.4 meV/atom higher in energy**, demonstrating that structural constraints imposed by the field introduce a significant thermodynamic penalty. Altogether, these results reveal that field-tailored synthesis drives $BaIrO_3$ into a compact and electronically unstable state, characterized by increased internal stress, charge redistribution, and enhanced bonding interactions, culminating in a metastable, high-energy configuration.

**Table 2** compares key structural, magnetic, and electronic properties of $BaIrO_3$ synthesized at ambient conditions and via magneto-synthesis, along with properties of ambient-grown samples measured under externally applied high pressure.

**Table 2: Comparison of $BaIrO_3$ Under Different Conditions**

| **Property/Characteristic** | **Ambient Synthesis** | **Magneto-Synthesis (< 0.1 T, M2)** | **High-Pressure Measurements [50]** |
|---|---|---|---|
| **Crystal Structure** | *Monoclinic (C2/m)* | *Monoclinic, compressed lattice* | *No phase transition up to 10 GPa* |
| **Ir1-Ir4 Bond Distance** | *~2.625 Å* | *2.605 Å, reduced by ~ 0.69% at 100 K* | *Slightly reduced (~ 4 GPa)* |
| **Unit Cell Volume V** | *848.10 Å$^3$ at 100 K* | *840.90 Å$^3$, reduced by 0.85% at 100 K* | *Reduced by ~ 0.80 % at ~ 3 GPa* |
| **Bond Angle Ir–O–Ir, θ** | *~155.5° at 100 K* | *~157.7° at 100 K* | – |
| **Magnetic Ordering $T_N$** | *185 K* | *Suppressed* | *Suppressed at ~ 4.5 GPa* |
| **Spin Structure** | *Canted AFM with weak ferromagnetism* | *Canted AFM suppressed* | *Retained under pressure* |
| **Magnetic Hysteresis** | *Strong* | *Nearly vanishes (suppressed canting)* | – |
| **Electrical Resistivity** | *Insulating* | *Drops by ~$10^4$; emergent metallic state* | *Remains insulating at 9 GPa* |
| **Specific Heat γ** | *γ ≈ 1.3 mJ/mol·K²* | *γ up to 35 mJ/mol·K² (correlated metal)* | – |
| **Metastability** | *Stable equilibrium phase* | *Metastable, high internal stress, ~ 18.4 meV/atom higher* | – |

*Notes: All values are taken from this work unless otherwise cited. The "High-Pressure Measurements" column refers to previous studies applying hydrostatic pressure [50].*

Given the dramatic changes observed across multiple properties, one might ask whether these effects could arise from impurity phases or disorder. The observed changes in structure and physical properties, spanning unit cell volume, Ir–Ir bond distances, Néel temperature, electrical resistivity, and specific heat, are both too large and too coherent to be attributed to extrinsic factors such as impurity phases or unintentional doping. This conclusion is also confirmed by our extensive x-ray diffraction (**Fig. 2** and **Supplementary Figs.1-2**) and SEM/EDX analyses (**Supplementary Information Section III**).

Electrical resistivity decreases by up to 4 orders of magnitude (**Fig. 3a**), the Sommerfeld coefficient increases nearly 10-fold (**Fig. 3d**), and the Ir–Ir bond length and unit cell volume $V$ contracts by as much as 0.69% and 0.85%, respectively (**Figs. 2a-2b**) and all in a manner that is consistent across multiple independently grown crystals and correlated with synthesis field strength **(Supplementary Information)**. **Such coherent and correlated trends are incompatible with dilute impurity effects, disorder, or off-stoichiometry which typically lead to localized, decoupled responses rather than coupled electronic, structural, and magnetic transformations**. Moreover, the field-tailored samples remain single-phase by x-ray diffraction and EDX and show no signs of secondary phases within the sensitivity of our measurements (**Supplementary Information**). First-principles calculations additionally reinforce that the field-stabilized phase lies significantly above the ground state in energy (**Fig. 4**), consistent with a metastable configuration, rather than impurity stabilization. All these results provide compelling evidence that the reported changes are intrinsic consequences of magneto-synthesis rather than extrinsic artifacts.

## IV. Conclusions

Our results demonstrate that magneto-synthesis via applying weak magnetic fields during high-temperature growth can dramatically alter the structural, magnetic, and electronic landscape of spin-orbit-coupled quantum materials. In $BaIrO_3$, we achieved structural compression, reduced lattice distortion, and bond-length modification that collectively suppressed antiferromagnetic order and enabled a striking insulator-to-metal transition. These changes are realized using fields $< 0.1$ T, which challenges the conventional belief that magnetic energies are negligible in chemical synthesis [24-25].

Our combined experimental and computational findings suggest that **magnetic fields can act as effective thermodynamic and kinetic levers**, stabilizing metastable and electronically unstable phases that conventional methods cannot access. As we advance toward high-field synthesis platforms such as our patented Field-Tailoring Technology (FTT) system with 5 T capabilities [53], the scope of magneto-synthesis is poised to expand significantly, unlocking new opportunities for the discovery and control of emergent quantum states. Beyond $BaIrO_3$, the broader implications of this work extend to a wide range of quantum materials where SOI, electronic correlations, or lattice instabilities play a central role [13, 26]. The ability to stabilize high-energy phases through modest field inputs may enable tailored control over phase competition, topological order, or symmetry breaking in high-Z materials. As such, magneto-synthesis could serve as a generalizable tool for engineering non-equilibrium states that are otherwise unattainable through traditional synthetic means.

**Methods**

Single crystals studied were grown using a flux method. The mixtures were fired at 1300 $^{o}$C for 15-20 hours and then slowly cooled to room temperature at a rate of 3 C/hour. Platinum crucibles were used. Measurements of crystal structures were performed using a Bruker Quest

ECO single-crystal diffractometer. It is equipped with an Oxford Cryosystem that creates sample temperature environments ranging from 80 K to 400 K during x-ray diffraction measurements. Chemical analyses of the samples were performed using a combination of a Hitachi MT3030 Plus Scanning Electron Microscope and an Oxford Energy Dispersive X-Ray Spectroscopy (EDX). Magnetic properties were measured using a Quantum Design (QD) MPMS-7 SQUID Magnetometer. Standard four-lead measurements of the electrical resistivity were carried out using a QD DynaCool PPMS System equipped with a 14-Tesla magnet. The heat capacity was measured down to 1.8 K using the PPMS.

First-principles calculations were performed using the Quantum ESPRESSO suite within the framework of density functional theory (DFT) [55]. The projector augmented wave (PAW) method was employed alongside the Perdew-Burke-Ernzerhof (PBE) generalized gradient approximation (GGA) for exchange-correlation effects. The kinetic energy cutoff for the charge density was set to 419.8 Ry, while the wavefunction cutoff was set to 46.6 Ry, ensuring accurate total energy convergence. Brillouin zone integrations were carried out using a $6 \times 9 \times 3$ *k*-mesh. To better capture the correlation effects in the Ir-5*d* orbitals, a Hubbard U correction of 2.0 eV was applied. pseudopotentials were sourced from Pslibrary [56]. Structural relaxation was conducted until the forces on atoms were below 4 eV/Å and the total stress was less than 5 kbar. Post processing bonding analysis was carried out using the LOBSTER (Local Orbital Basis Suite Towards Electronic-Structure Reconstruction) code [57, 58], which projects the plane-wave-based DFT wavefunctions onto a local atomic basis. This enabled the extraction of Mulliken gross populations, ICOHP, and ICOBI values, providing a detailed view of charge distribution, bonding energy, and bond order across different structural states.

## ACKNOWLEDGEMENTS

This work was supported by the US DOE via Award DE-SC0025273. X.H. and T.T.T thank the Arnold and Mabel Beckman Foundation, the NSF Award NSF-DMR-2338014, and the Camille Henry Dreyfus Foundation.

**FIGURE CAPTIONS**

**Fig. 1. Magneto-Synthesis and lattice modification in $BaIrO_3$: a, Schematic of** field-assisted crystal growth setup using permanent magnets placed right outside of the furnace to apply weak magnetic fields (< 0.1 T) during synthesis. **b, Furnace-floor magnetic field distributions** for M1 and M2 and approximate locations (white circles) of crucibles relative to the magnet(s). **c, Trimer** consisting of three face-sharing $IrO_6$ octahedra; the Ir1-Ir4 bond distance $d_{Ir1\text{-}Ir4}$ is crtical to the ground state. **d, Crystal structure in the ac planes**, showing corner-sharing $IrO_6$ octahedra. The Ir-O-Ir bond angle θ is a key structural parameter that controls electronic and magnetic properties. The two red arrows schematically illustrate the unit cell compression due to magneto-synthesis.

Note that the *a* and *c* axes shrink while the *b* axis expands slightly. **e, Empirical phase diagram for high-Z trimer lattices** correlating magnetic ordering temperature $T_N$ with average metal-metal bond distance, a unique tunning knob for controlling the ground state of trimer lattices. Long-range order collapses below ~2.58 Å.

**Fig. 2. Structural contraction and suppression of magnetic order in field-tailored $BaIrO_3$: a-b, Temperature dependence** of the Ir1-Ir4 bond distance (**a**) and unit cell volume V (**b**) for non-tailored and field-tailored samples (M1, M2), showing up to 0.69% bond contraction and 0.85% volume reduction via magneto-synthesis. **c, Representative crystal. d, Magnetic structure** of non-tailored $BaIrO_3$ determined from neutron diffraction. Spins align along the *c* axis with a canted configuration within trimers. **e, Magnetic susceptibility for $BaIrO_3$:** The temperature dependence of the c-axis magnetic susceptibility $\chi_c$ showing suppressions of the transition temperature $T_N$ from 185 K (non-tailored) to progressively lower values in M1, M2 and M2* field-tailored samples. **f, Isothermal magnetization M for $BaIrO_3$ at 5 K.** The canted spin structure in the non-tailored $BaIrO_3$ leads to domain-like behavior, and thus strong magnetic hysteresis in M in which $M_a > M_c$. **g,** Field-tailored M2 sample shows nearly vanishing hysteresis, indicating reduced spin canting and domain activity due to structural relaxation.

**Fig. 3. Emergent metallicity and enhanced correlations induced by magneto-synthesis:** Temperature dependence of *c*-axis resistivity $\rho_c$ for non-tailored and field-tailored samples. Field-tailored sample M2 exhibits a four-order-of-magnitude drop in $\rho_c$, indicating an insulator-to-metal transition. **b, Magnetic field dependence of the *c*-axis resistivity $\rho_c$** at T = 2 K for H || *a* axis. Non-tailored $BaIrO_3$ shows large positive magnetoresistance, while field-tailored M2 remains nearly field-independent, consistent with metallic behavior. **c, The temperature dependence of the specific heat capacity C(T) and corresponding entropy removal ΔS (right scale)**; the field-

tailored sample M1 exhibits a large anomaly at $T_N$, suggesting a highly ordered magnetic structure.

**d, Low-temperature specific heat** plotted as C/T vs $T^2$. Field-tailored samples exhibit enhanced Sommerfeld coefficients $\gamma$, reflecting increased density of states at the Fermi level.

**Fig. 4. First-principles analysis of structural metastability and electronic instability in BaIrO:** **a,** Unit cell volume and lattice parameters *a*, *b*, and *c* of the five structures. **b,** Internal force and stress magnitudes, highlighting increasing mechanical strain from relaxation to field tailoring. **c,** Mulliken charge populations on Ir-5*d* and Ba-6*s* orbitals, illustrating charge redistribution under constrained geometry. **d,** Relative total energy per atom with respect to the fully relaxed structure, quantifying the thermodynamic cost of each structural configuration; and band energy showing cumulative electronic destabilization near the Fermi level.

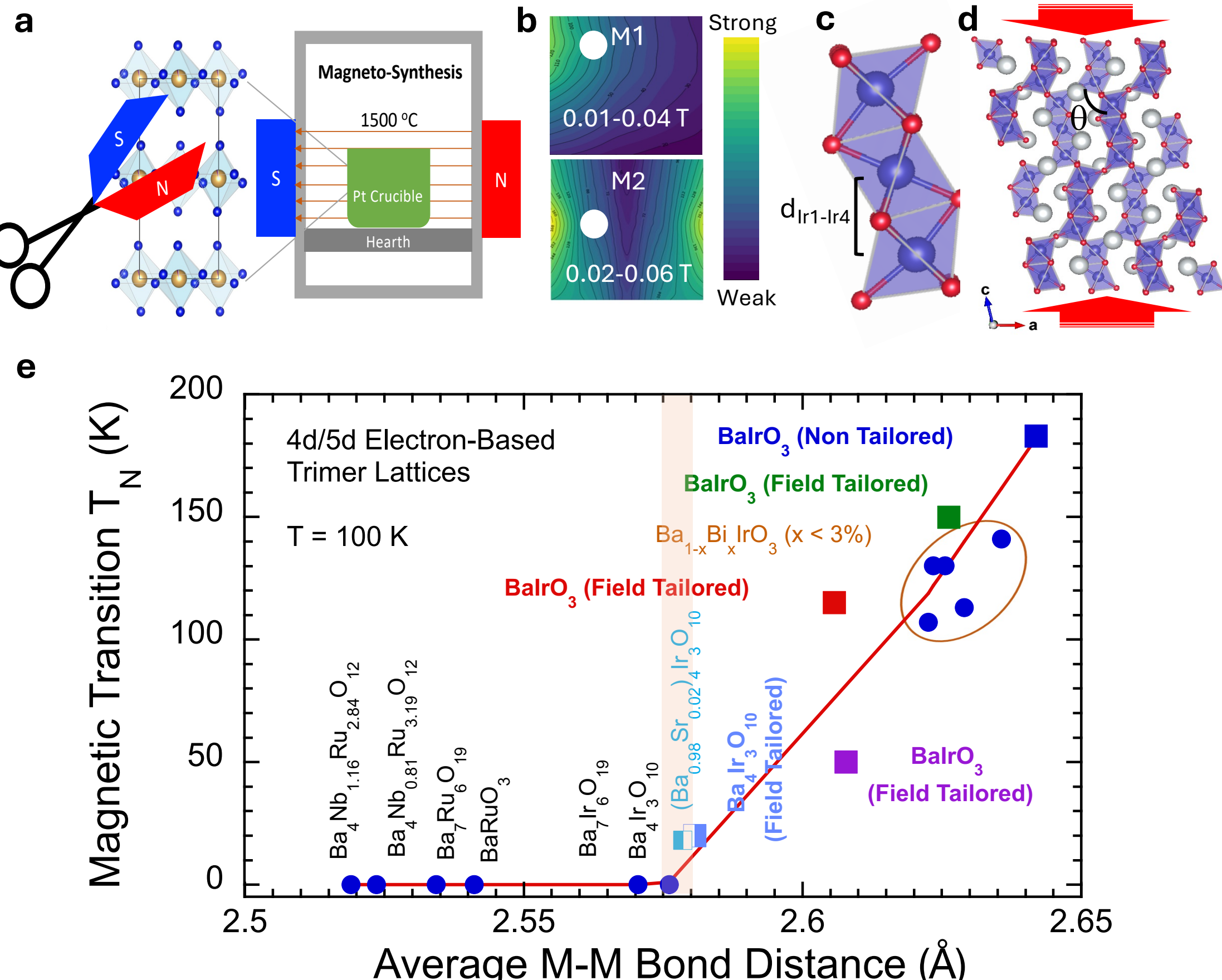

Figure 1

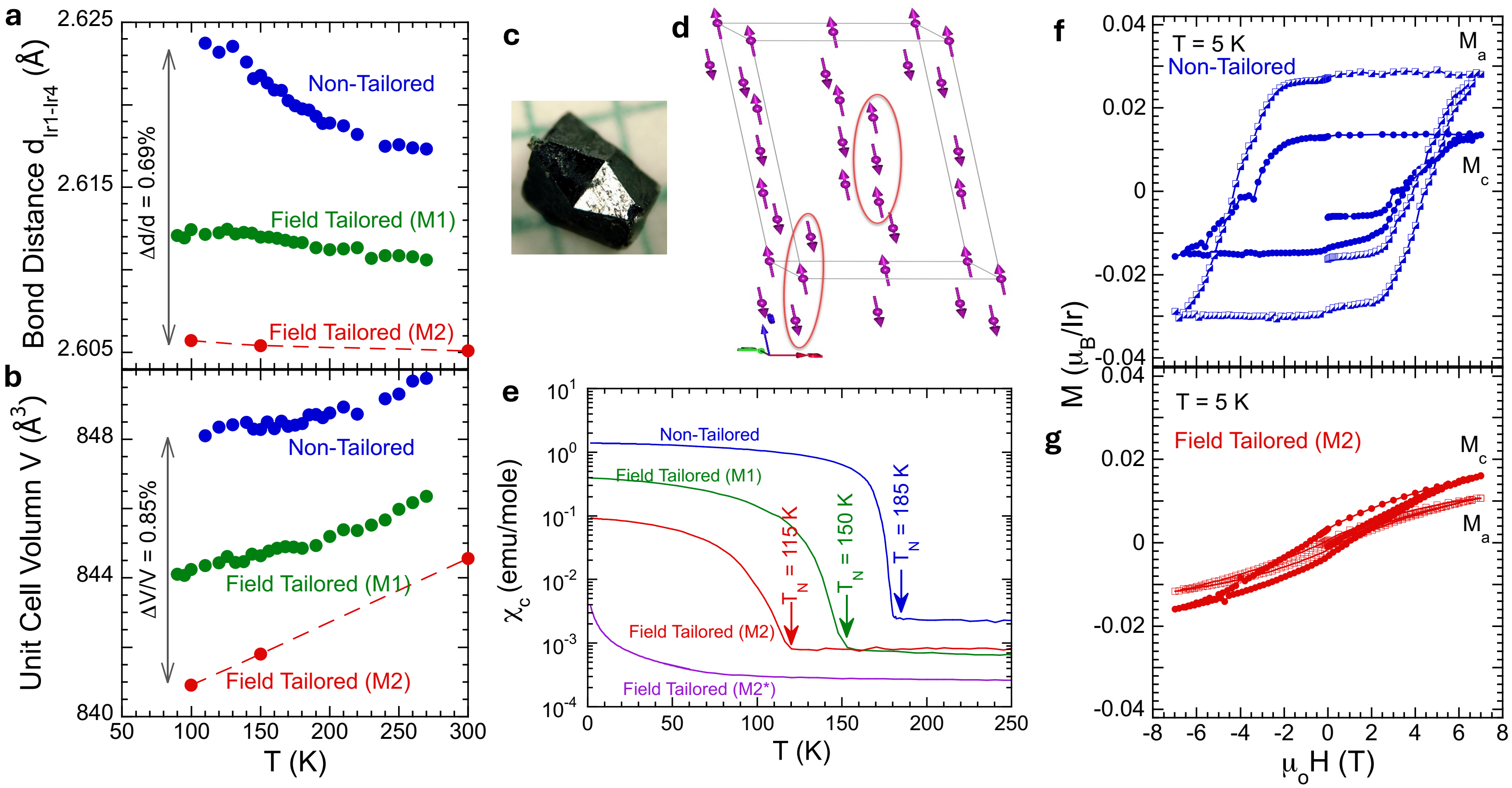

Figure 2

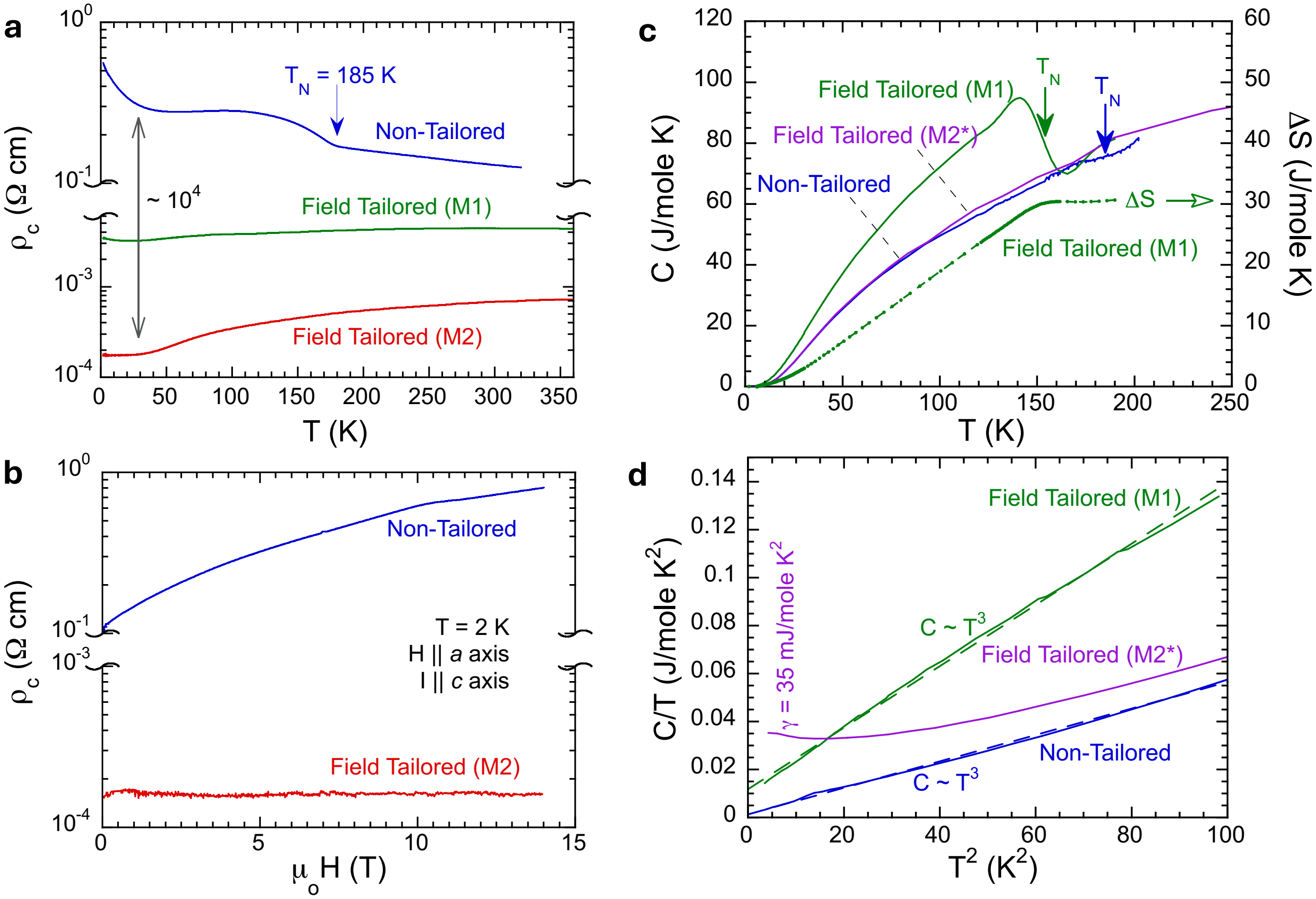

Figure 3

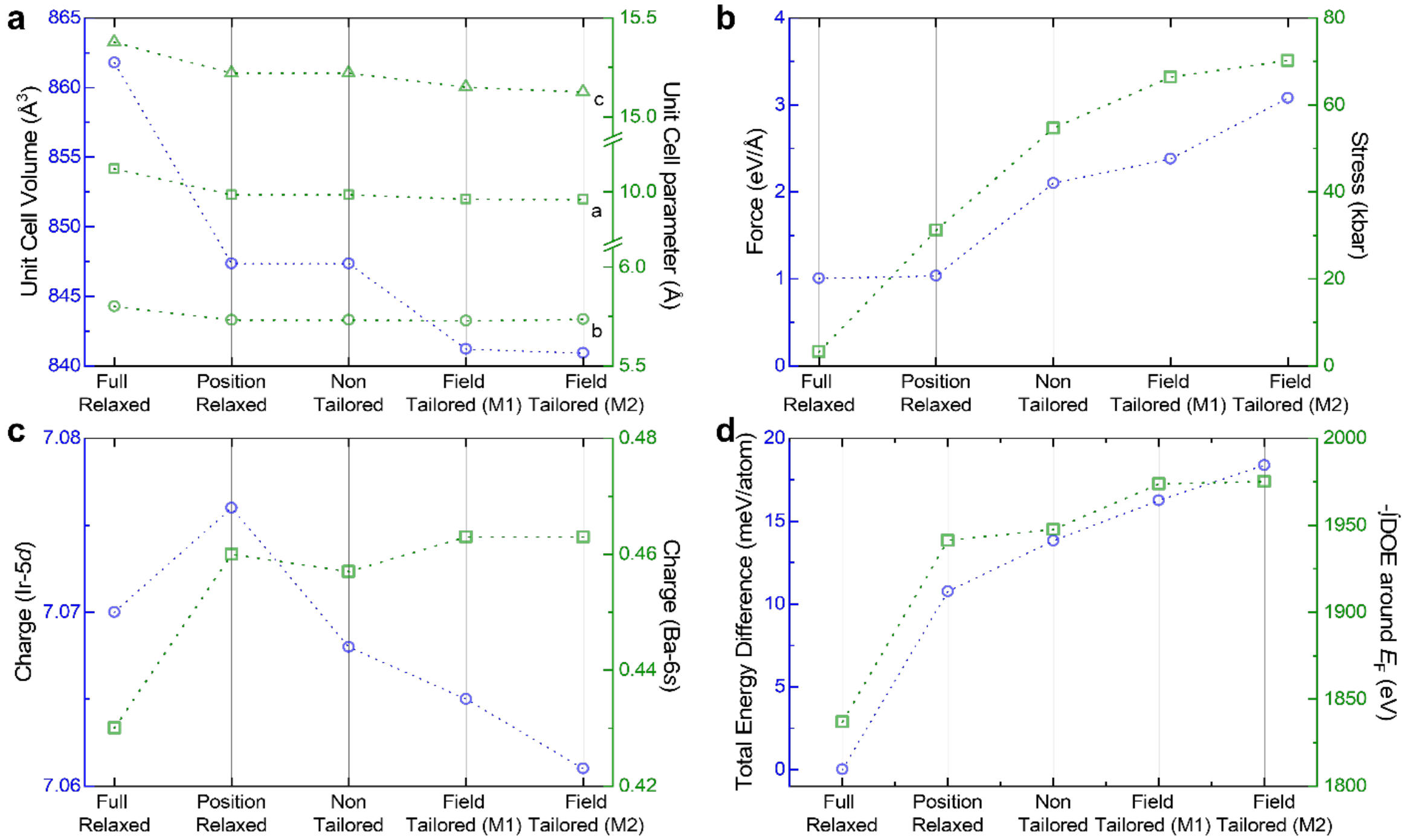

Figure 4

Supplementary Information

# Field-Tailoring Quantum Materials via Magneto-Synthesis: Metastable Metallic and Magnetically Suppressed Phases in a Trimer Iridate

Tristan R. Cao[1], Hengdi Zhao[1], Xudong Huai[2], Arabella Quane[1], Thao T. Tran[2], Feng Ye[3] and Gang Cao[1]

[1]Department of Physics, University of Colorado at Boulder, Boulder, CO 80309

[2]Department of Chemistry, Clemson University, Clemson, SC 29634

[3]Oak Ridge National Laboratory, Oak Ridge, TN 37830

## I. Experimental Details

Single crystals studied were grown using a flux method. The mixtures were fired at 1300 $^{o}$C for 15-20 hours and then slowly cooled to room temperature at a rate of 3 C/hour. Platinum crucibles were used. Measurements of crystal structures were performed using a Bruker Quest ECO single-crystal diffractometer. It is equipped with an Oxford Cryosystem that creates sample temperature environments ranging from 80 K to 400 K during x-ray diffraction measurements. Chemical analyses of the samples were performed using a combination of a Hitachi MT3030 Plus Scanning Electron Microscope and an Oxford Energy Dispersive X-Ray Spectroscopy (EDX). Magnetic properties were measured using a Quantum Design (QD) MPMS-7 SQUID Magnetometer. Standard four-lead measurements of the electrical resistivity were carried out using a QD DynaCool PPMS System equipped with a 14-Tesla magnet. The heat capacity was measured down to 1.8 K using the PPMS.

First-principles calculations were performed using the Quantum ESPRESSO suite within the framework of density functional theory (DFT) [1]. The projector augmented wave (PAW) method was employed alongside the Perdew-Burke-Ernzerhof (PBE) generalized gradient approximation (GGA) for exchange-correlation effects. The kinetic energy cutoff for the charge density was set to 419.8 Ry, while the wavefunction cutoff was set to 46.6 Ry, ensuring accurate total energy convergence. Brillouin zone integrations were carried out using a $6 \times 9 \times 3$ *k*-mesh. To better capture the correlation effects in the Ir-5*d* orbitals, a Hubbard U correction of 2.0 eV was applied. pseudopotentials were sourced from Pslibrary [2]. Structural relaxation was conducted until the forces on atoms were below 4 eV/Å and the total stress was less than 5 kbar. Post processing bonding analysis was carried out using the LOBSTER (Local Orbital Basis Suite Towards Electronic-Structure Reconstruction) code [3, 4], which projects the plane-wave-based DFT wavefunctions onto a local atomic basis. This enabled the extraction of Mulliken gross populations, ICOHP, and ICOBI values, providing a detailed view of charge distribution, bonding energy, and bond order across different structural states.

## II. Supplementary Figures for Lattice Parameters and DFT Calculations

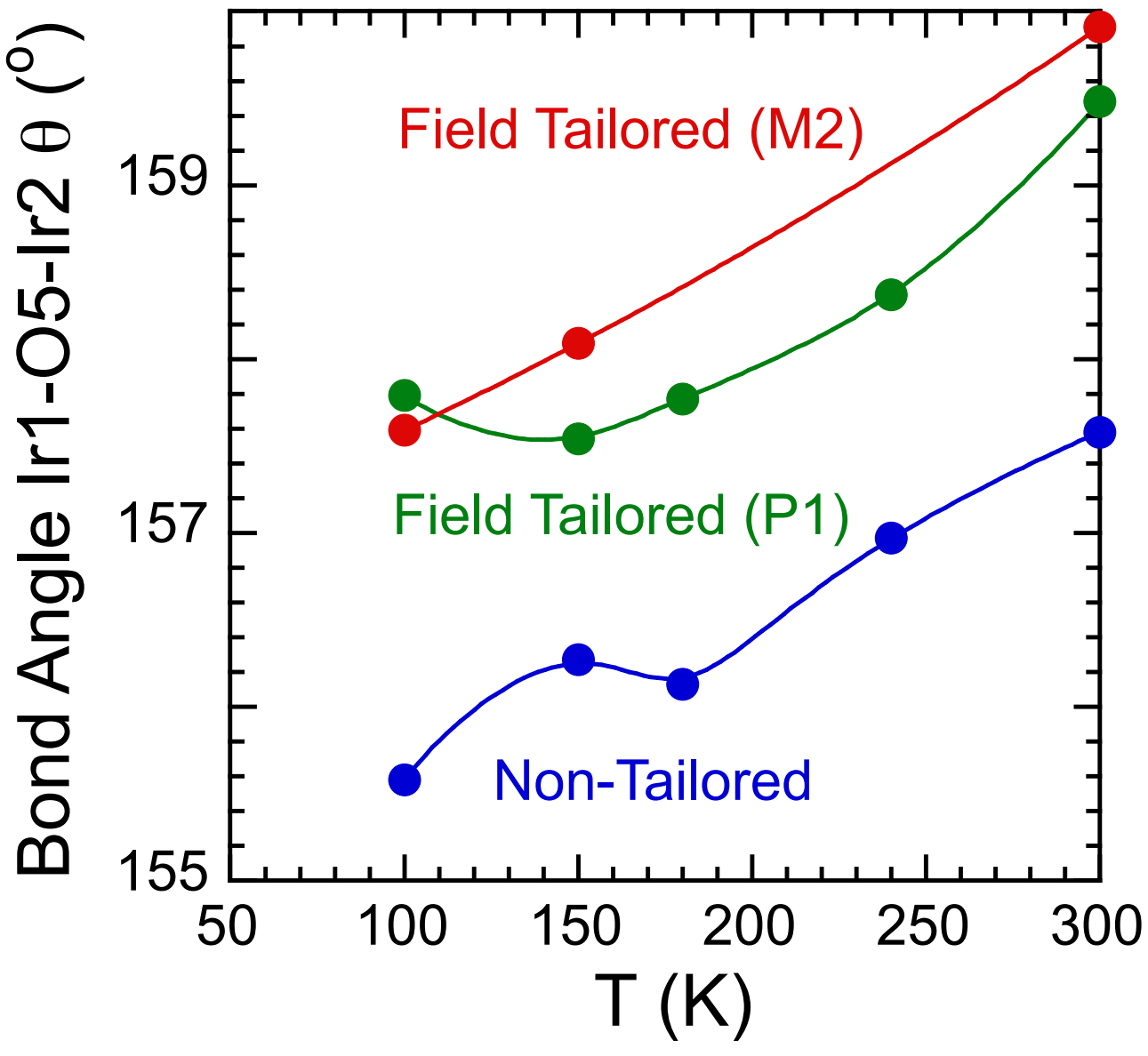

**Supplementary Fig. 1.** Bond angle Ir1-O5-Ir2 as a function of temperature

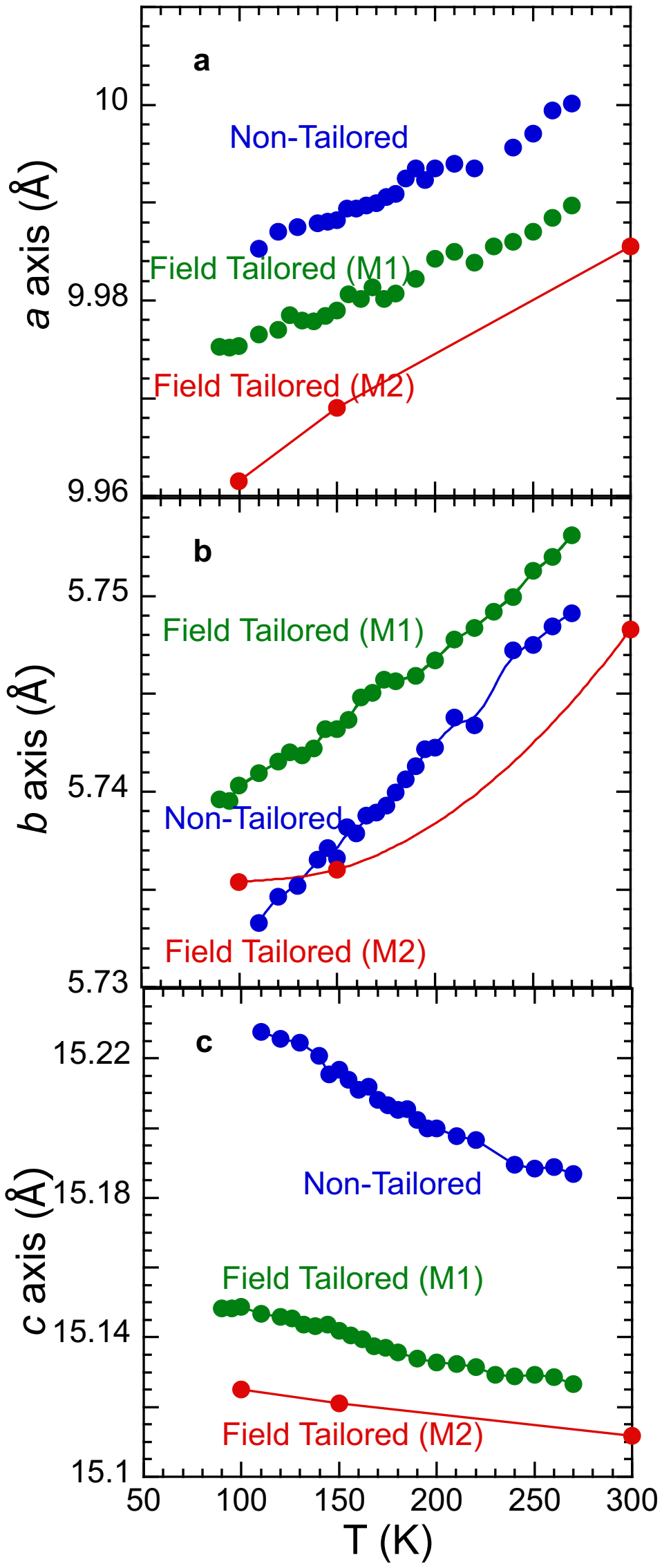

**Supplementary Fig. 2.** Temperature dependence of the lattice parameters (a) *a* axis, (b) *b* axis, and (c) *c* axis.

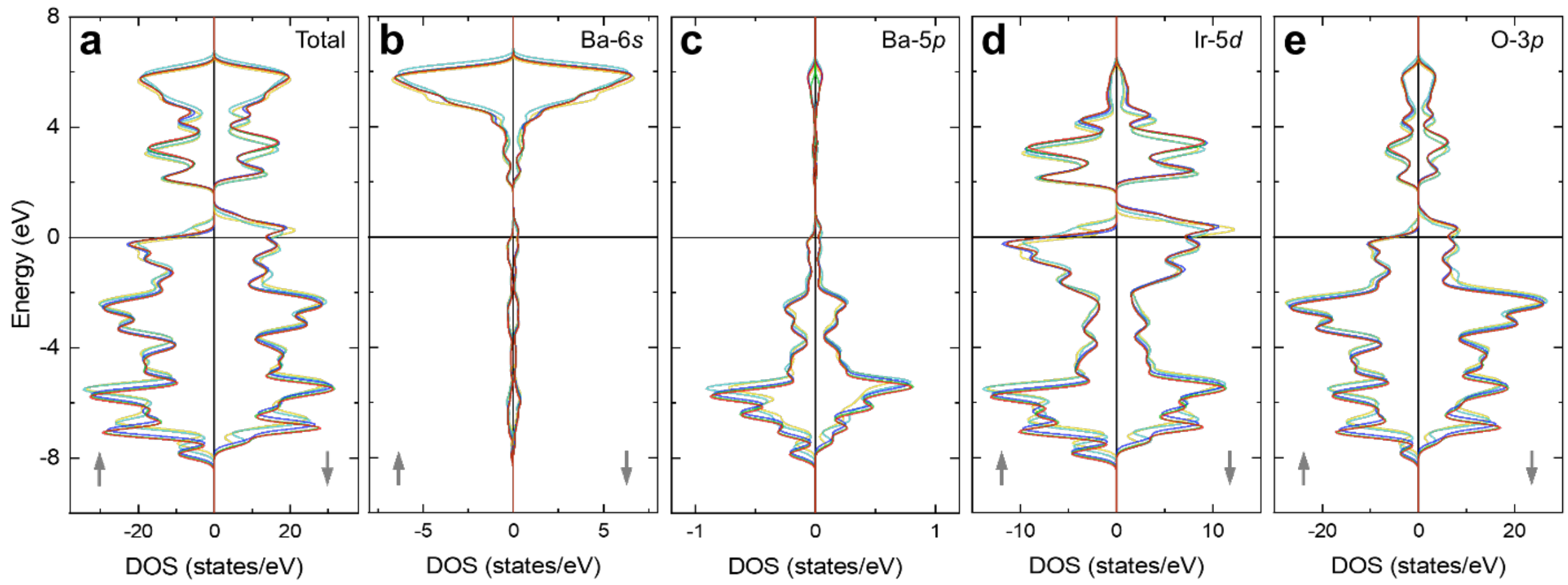

**Supplementary Fig. 3.** DOS of five structures: fully relaxed (yellow), position-relaxed (cyan), non-tailored (blue), field-tailored (M1) (green), and field-tailored (M2) (red).

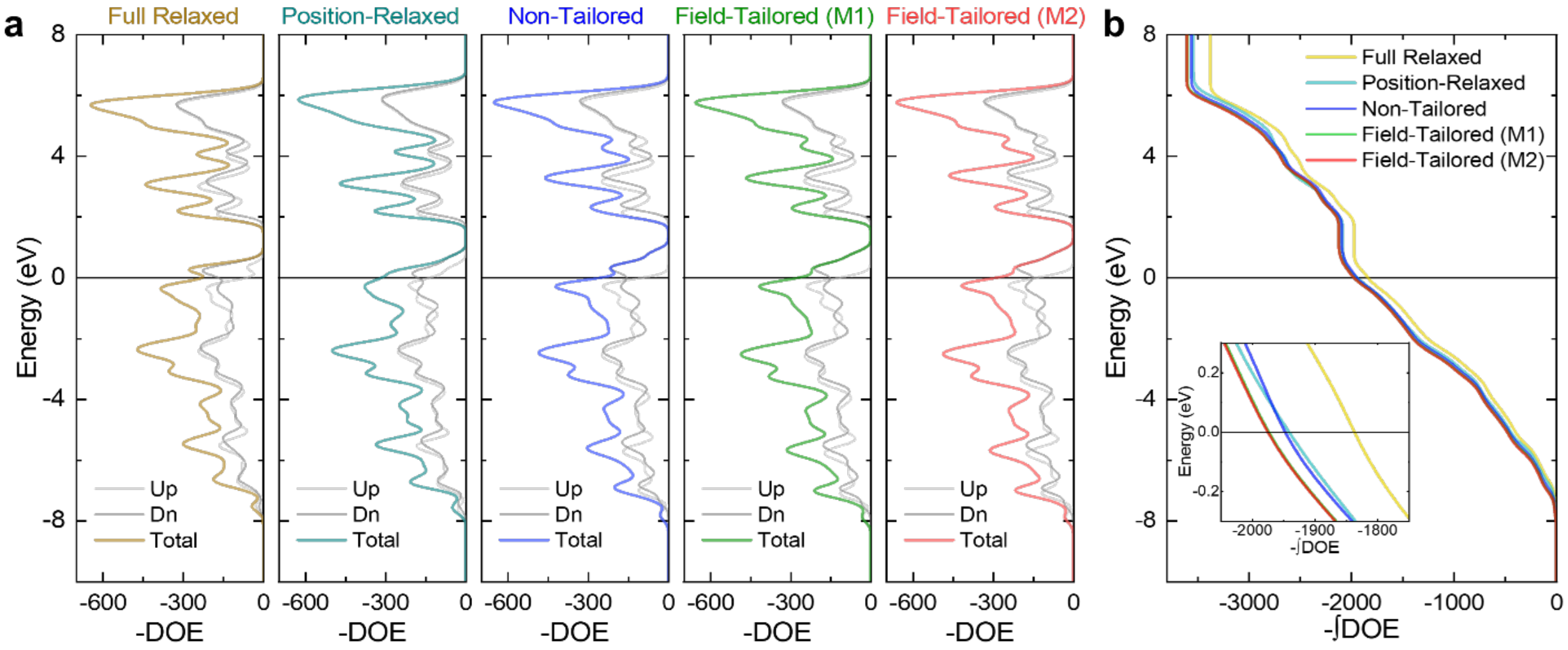

**Supplementary Fig 4.** (a) DOE of five structures: fully relaxed, position-relaxed, non-tailored, field-tailored (M1), and field-tailored (M2). (b) Integrated density of energy (band energy) at around Fermi level.

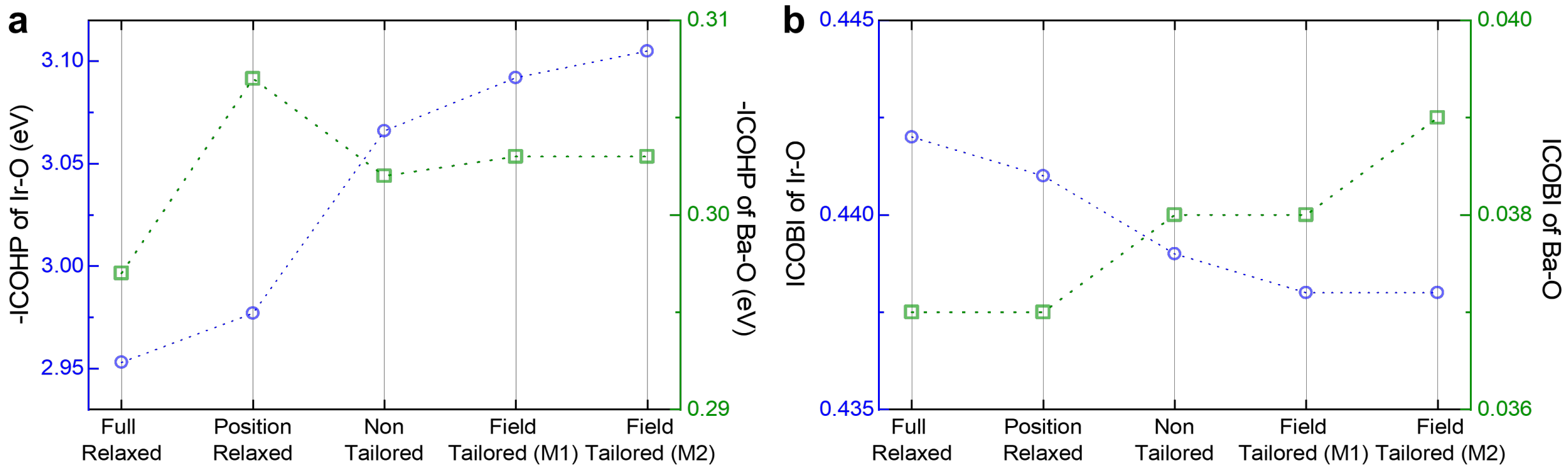

**Supplementary Fig. 5.** Integrated crystal orbital Hamilton population (-ICOHP) showing increases for both Ir-O and Ba-O bonds in the field-tailored structures. (b) Integrated crystal orbital bond index (ICOBI), revealing a covalent-like character for Ir-O bonds and ionic nature for Ba-O bonds, the distinct trends are in good agreement with the charge redistribution.

## III. Supplementary Data on EDX and X-Ray Diffraction Analysis and Stoichiometry Confirmation

The following figures present representative energy-dispersive x-ray spectroscopy (EDX) data collected from non-tailored ($T_N$ = 185 K), field-tailored M1 ($T_N$ =150) and M2* ($T_N$ ~ 0) $BaIrO_3$ single crystals. These measurements serve to confirm that the elemental ratio between Ba and Ir remains consistent across all samples, irrespective of applied field conditions during synthesis. No evidence of extrinsic impurity phases was detected within the resolution limits of the technique.

It is important to note that EDX analysis of $BaIrO_3$ can be complicated by **spectral peak overlap** between Ba and Ir. Specifically, the L-series x-ray emission lines of Ba and the M-series lines of Ir are close in energy and partially overlap. As a result, automated peak deconvolution software may incorrectly assign intensity from the Ir-M signal to the Ba-L peak, artificially elevating the apparent Ba content. This artifact is well-documented in materials containing both Ba and Ir and should not be interpreted as stoichiometric deviation or impurity incorporation.

Despite this limitation, the combined use of EDX and single-crystal x-ray diffraction provides a reliable picture: x-ray data (see **Main Text Fig. 2**, **Supplementary Figs. 1-2** and **Tables S1, S1 Extended, S2 and S2 Extended below** for representative samples) confirm single-phase crystallinity with no superstructure or secondary peaks. All these results provide strong evidence that the dramatic changes in physical properties observed in the field-tailored samples are not driven by off-stoichiometry or contamination, but instead reflect intrinsic, bulk modifications induced by magneto-synthesis.

**A. EDX Spectra for Non-Tailored and Field Tailored Samples**

**Non-Tailored Sample ($T_N$ ~ 185 K)**

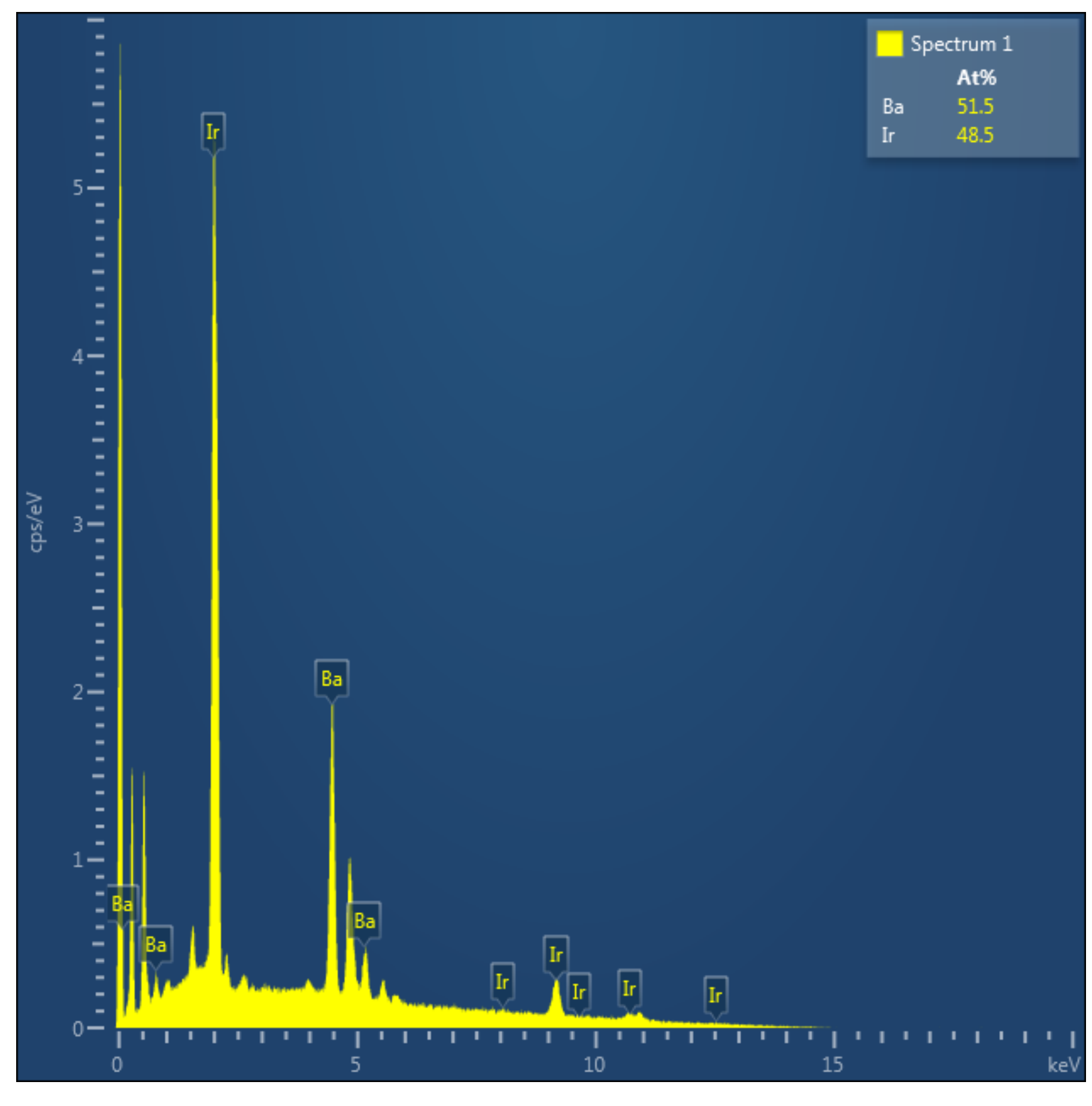

**Field Tailored Sample M2* ($T_N \sim 0$ K)**

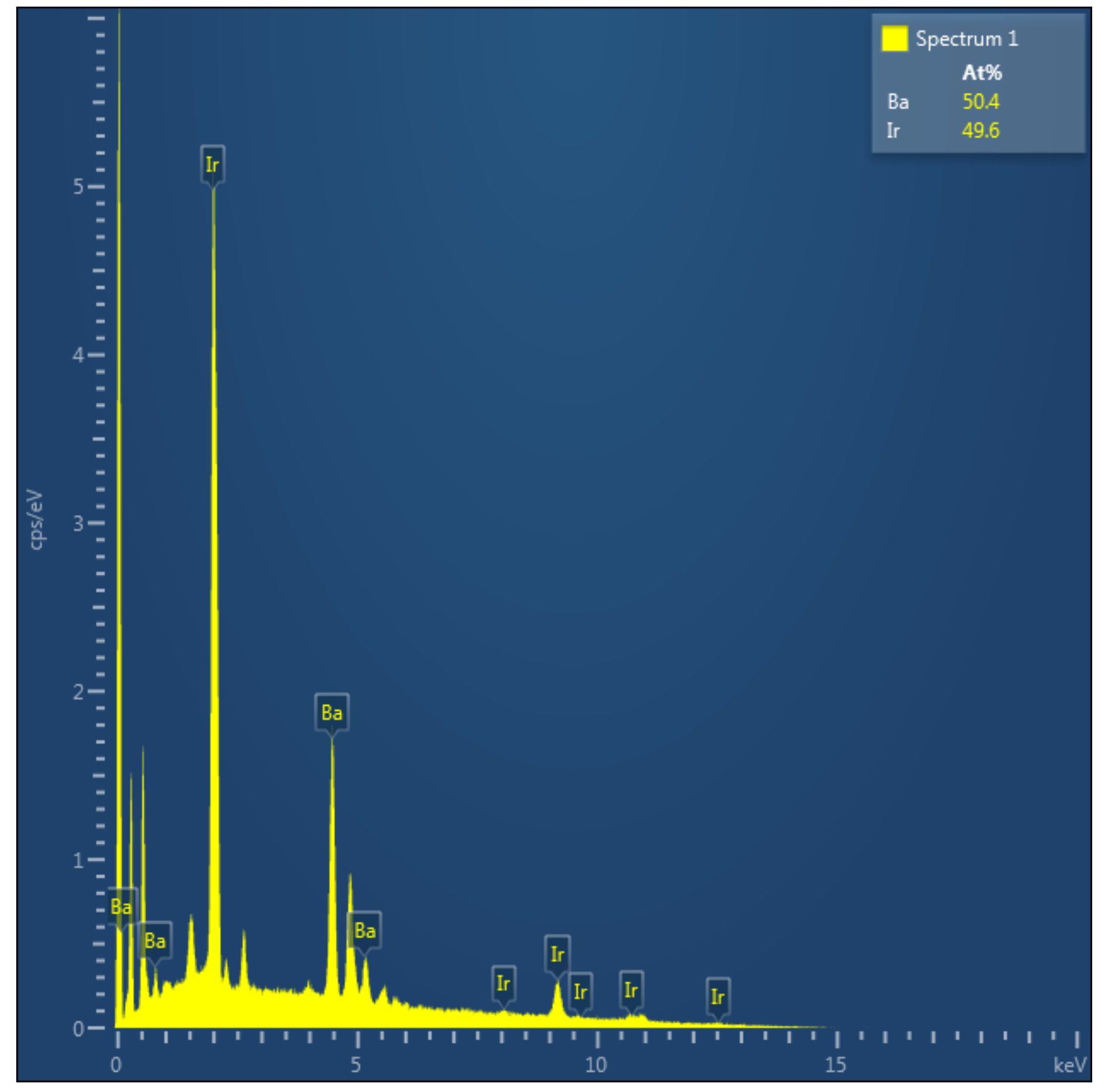

**Field Tailored Sample M1 ($T_N$ ~ 150 K)**

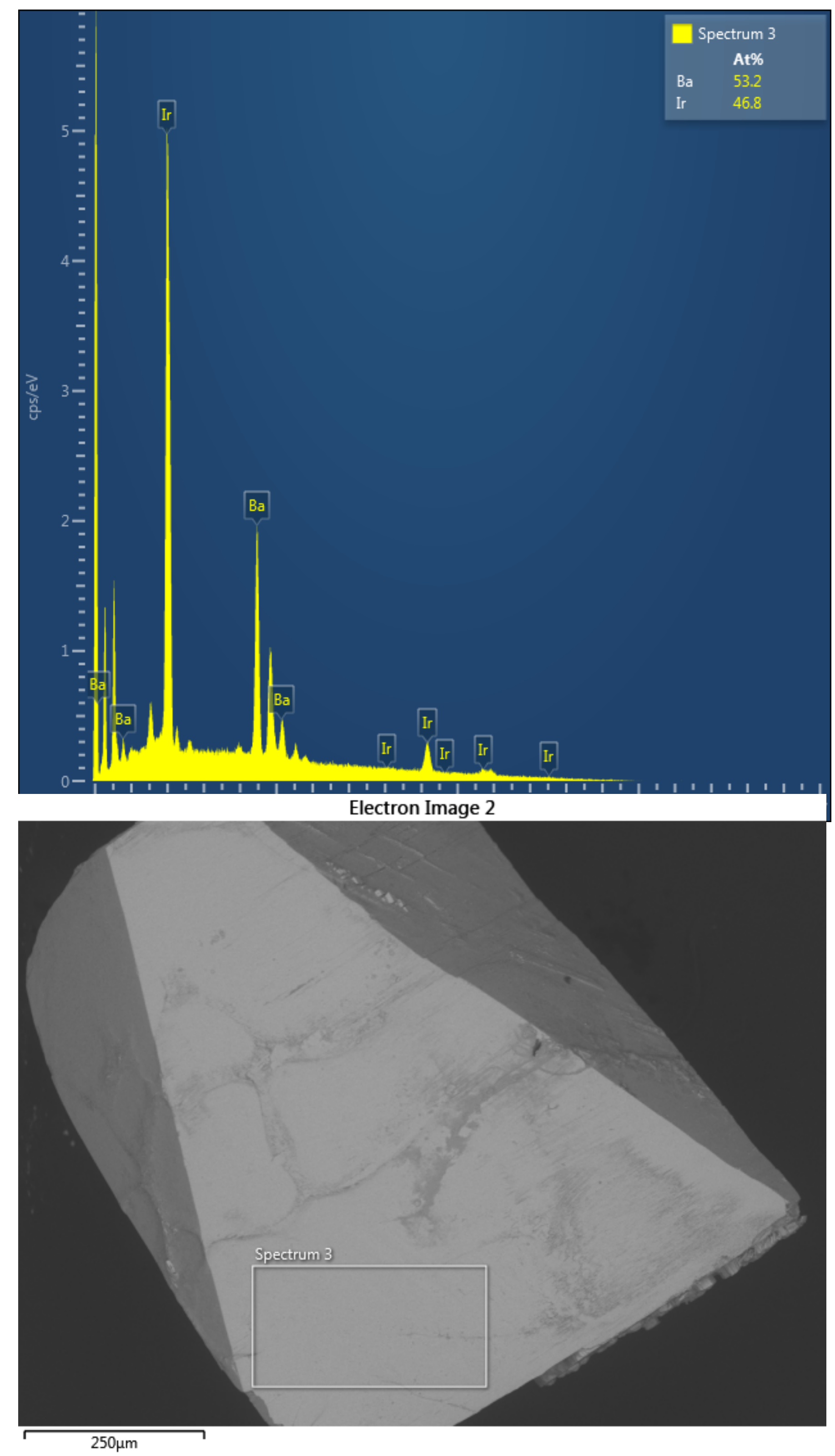

**B. Tables for X-Ray Diffraction Details for Non-Tailored and Field Tailored Samples**

**Table S1. Single crystal diffraction results of Batch #353 (M1)** in the space group *C*2/*m* at 100 K.

| Empirical formula | $BaIrO_3$ |
|---|---|
| Formula weight | 377.54 |
| Temperature/K | 99.99 |
| Crystal system | monoclinic |
| Space group | C2/m |
| a/Å | 9.9632(4) |
| b/Å | 5.7272(2) |
| c/Å | 15.1501(6) |
| α/° | 90 |
| β/° | 103.3269(14) |
| γ/° | 90 |
| Volume/Å$^3$ | 841.20(6) |
| Z | 12 |
| $\rho_{calc}$g/cm$^3$ | 8.943 |
| μ/mm$^{-1}$ | 61.123 |
| F(000) | 1884.0 |
| Radiation | MoKα (λ = 0.71073) |
| 2Θ range for data collection/° | 5.526 to 80.806 |
| Index ranges | $-18 \leq h \leq 18$, $-10 \leq k \leq 10$, $-27 \leq l \leq 27$ |
| Reflections collected | 65867 |
| Independent reflections | 2876 [$R_{int}$ = 0.0525, $R_{sigma}$ = 0.0159] |
| Data/restraints/parameters | 2876/0/85 |
| Goodness-of-fit on $F^2$ | 1.203 |
| Final R indexes [I>=2σ (I)] | $R_1$ = 0.0294, $wR_2$ = 0.0673 |
| Final R indexes [all data] | $R_1$ = 0.0320, $wR_2$ = 0.0681 |
| Largest diff. peak/hole / e Å$^{-3}$ | 6.81/-5.23 |

$R = \Sigma||F_o|-|F_c||/\Sigma|F_o|$, $wR = [\Sigma[w(|F_o|^2-|F_c|^2)^2]/\Sigma[w(|F_o|^4)]]^{1/2}$ and $w=1/(\sigma^2(I)+0.0016I^2)$

**Table S1 Extended** Fractional Atomic Coordinates (×10$^4$) and Equivalent Isotropic Displacement Parameters (Å$^2$×10$^3$) for Batch 353 (M1). $U_{eq}$ is defined as 1/3 of the trace of the orthogonalized $U_{IJ}$ tensor.

| **Atom** | ***x*** | ***y*** | ***z*** | **U(eq)** |
|---|---|---|---|---|
| Ir1 | 4137.1(3) | 0 | 3236.8(2) | 5.41(6) |
| Ir2 | 10311.4(3) | 0 | 1779.5(2) | 5.47(6) |

| **Atom** | ***x*** | ***y*** | ***z*** | **U(eq)** |
|---|---|---|---|---|
| Ir3 | 10000 | 0 | 0 | 4.46(7) |
| Ir4 | 5000 | 0 | 5000 | 4.56(7) |
| Ba5 | 7209.8(5) | 0 | 2514.1(3) | 6.36(8) |
| Ba6 | 3497.9(5) | 0 | 775.8(3) | 6.22(8) |
| Ba7 | 1314.5(5) | 0 | 4274.1(3) | 6.97(8) |
| O1 | 10962(4) | 2288(8) | 966(3) | 6.2(6) |
| O2 | 8591(6) | 0 | 765(4) | 6.6(9) |
| O3 | 3830(4) | -2352(8) | 4165(3) | 6.7(6) |
| O4 | 6086(6) | 0 | 4016(4) | 6.2(9) |
| O5 | 12071(7) | 0 | 2720(5) | 8.8(10) |
| O6 | 9456(5) | -2547(9) | 2376(3) | 9.8(7) |

**Table S1 Extended** Anisotropic Displacement Parameters ($Å^2 \times 10^3$) for Batch 353 (M1). The Anisotropic displacement factor exponent takes the form: $-2\pi^2[h^2a^{*2}U_{11}+2hka^*b^*U_{12}+\ldots]$.

| **Atom** | **$U_{11}$** | **$U_{22}$** | **$U_{33}$** | **$U_{23}$** | **$U_{13}$** | **$U_{12}$** |
|---|---|---|---|---|---|---|
| Ir1 | 5.41(11) | 4.69(11) | 6.08(11) | 0 | 1.26(8) | 0 |
| Ir2 | 5.32(11) | 4.90(11) | 6.00(11) | 0 | 0.92(8) | 0 |
| Ir3 | 3.94(14) | 4.00(14) | 5.34(14) | 0 | 0.88(10) | 0 |
| Ir4 | 4.21(14) | 3.77(14) | 5.76(14) | 0 | 1.26(11) | 0 |
| Ba5 | 5.65(16) | 4.75(16) | 8.67(17) | 0 | 1.66(12) | 0 |
| Ba6 | 5.34(16) | 4.65(16) | 7.91(17) | 0 | -0.03(12) | 0 |
| Ba7 | 6.89(17) | 4.89(17) | 9.96(18) | 0 | 3.64(13) | 0 |
| O1 | 6.9(15) | 5.0(15) | 6.1(15) | 0.3(12) | 0.5(11) | -1.5(12) |
| O2 | 3(2) | 10(2) | 7(2) | 0 | 2.1(16) | 0 |
| O3 | 6.6(15) | 5.5(16) | 7.6(15) | -0.2(13) | 0.7(11) | -0.8(12) |
| O4 | 2.6(19) | 9(2) | 8(2) | 0 | 1.9(16) | 0 |
| O5 | 9(2) | 10(3) | 7(2) | 0 | 0.4(18) | 0 |
| O6 | 12.3(18) | 6.6(17) | 11.8(18) | 3.7(15) | 5.5(14) | -0.3(14) |

**Table S1 Extended** Bond Lengths for Batch 353 (M1)

| | | | | | |
|---|---|---|---|---|---|
| Ir1 | Ir4 | 2.6107(3) | Ir4 | O3[16] | 2.024(4) |
| Ir1 | Ba5[1] | 3.4810(3) | Ir4 | O3[3] | 2.024(4) |
| Ir1 | Ba5 | 3.4808(6) | Ir4 | O3 | 2.024(5) |
| Ir1 | Ba5[2] | 3.4810(3) | Ir4 | O4 | 2.034(6) |
| Ir1 | Ba6 | 3.6348(6) | Ir4 | O4[15] | 2.034(6) |
| Ir1 | Ba7 | 3.5236(6) | Ba5 | O1[1] | 2.848(4) |
| Ir1 | O3[3] | 2.021(5) | Ba5 | O1[17] | 2.848(4) |
| Ir1 | O3 | 2.021(5) | Ba5 | O2 | 3.256(6) |

| Ir1 | O4 | 2.027(6) | Ba5 | O3$^{18}$ | 3.049(4) |
|---|---|---|---|---|---|
| Ir1 | O5$^{4}$ | 2.027(7) | Ba5 | O3$^{7}$ | 3.049(4) |
| Ir1 | O6$^{2}$ | 1.991(5) | Ba5 | O4 | 2.758(6) |
| Ir1 | O6$^{5}$ | 1.991(5) | Ba5 | O5$^{1}$ | 2.8874(9) |
| Ir2 | Ir3 | 2.6418(3) | Ba5 | O5$^{2}$ | 2.8874(9) |
| Ir2 | Ba5$^{6}$ | 3.4707(3) | Ba5 | O6 | 2.719(5) |
| Ir2 | Ba5 | 3.5175(6) | Ba5 | O6$^{3}$ | 2.719(5) |
| Ir2 | Ba5$^{7}$ | 3.4707(3) | Ba5 | O6$^{5}$ | 3.047(5) |
| Ir2 | Ba6$^{7}$ | 3.5370(3) | Ba5 | O6$^{2}$ | 3.047(5) |
| Ir2 | Ba6$^{6}$ | 3.5370(3) | Ba6 | O1$^{1}$ | 2.863(5) |
| Ir2 | O1 | 2.006(4) | Ba6 | O1$^{19}$ | 3.213(4) |
| Ir2 | O1$^{3}$ | 2.006(4) | Ba6 | O1$^{4}$ | 2.919(4) |
| Ir2 | O2 | 2.021(6) | Ba6 | O1$^{10}$ | 3.213(4) |
| Ir2 | O5 | 1.988(7) | Ba6 | O1$^{17}$ | 2.863(5) |
| Ir2 | O6 | 2.007(5) | Ba6 | O1$^{20}$ | 2.919(4) |
| Ir2 | O6$^{3}$ | 2.007(5) | Ba6 | O2$^{1}$ | 2.8653(2) |
| Ir3 | Ba6$^{8}$ | 3.4116(5) | Ba6 | O2$^{8}$ | 2.746(6) |
| Ir3 | Ba6$^{6}$ | 3.5542(3) | Ba6 | O2$^{2}$ | 2.8653(2) |
| Ir3 | Ba6$^{9}$ | 3.4116(5) | Ba6 | O6$^{2}$ | 2.774(5) |
| Ir3 | Ba6$^{10}$ | 3.5542(3) | Ba6 | O6$^{5}$ | 2.774(5) |
| Ir3 | O1$^{11}$ | 2.033(4) | Ba7 | O3$^{5}$ | 2.875(5) |
| Ir3 | O1$^{12}$ | 2.033(4) | Ba7 | O3$^{13}$ | 2.841(5) |
| Ir3 | O1$^{3}$ | 2.033(4) | Ba7 | O3$^{21}$ | 2.841(5) |
| Ir3 | O1 | 2.033(4) | Ba7 | O3$^{2}$ | 2.875(5) |
| Ir3 | O2 | 2.016(6) | Ba7 | O3 | 2.883(4) |
| Ir3 | O2$^{11}$ | 2.015(6) | Ba7 | O3$^{3}$ | 2.883(5) |
| Ir4 | Ba7$^{13}$ | 3.4335(3) | Ba7 | O4$^{15}$ | 3.213(6) |
| Ir4 | Ba7$^{14}$ | 3.4335(3) | Ba7 | O4$^{1}$ | 2.8921(9) |
| Ir4 | Ba7$^{6}$ | 3.4335(3) | Ba7 | O4$^{2}$ | 2.8921(9) |
| Ir4 | Ba7$^{7}$ | 3.4335(3) | Ba7 | O5$^{4}$ | 2.632(7) |
| Ir4 | O3$^{15}$ | 2.024(5) | | | |

**Table S2. Single crystal diffraction results of Batch 916 (Non-Tailored)** in the space group *C2/m* at 100 K.

| Empirical formula | $BaIrO_3$ |
|---|---|
| Formula weight | 377.54 |
| Temperature/K | 99.65 |
| Crystal system | monoclinic |
| Space group | C2/m |
| a/Å | 9.9855(8) |
| b/Å | 5.7314(5) |
| c/Å | 15.2203(13) |
| α/° | 90 |
| β/° | 103.403(4) |
| γ/° | 90 |
| Volume/Å$^3$ | 847.35(12) |
| Z | 12 |
| $\rho_{calc}$g/cm$^3$ | 8.878 |
| μ/mm$^{-1}$ | 60.680 |
| F(000) | 1884.0 |
| Radiation | MoKα (λ = 0.71073) |
| 2Θ range for data collection/° | 5.502 to 80.862 |
| Index ranges | $-18 \leq h \leq 18$, $-10 \leq k \leq 10$, $-27 \leq l \leq 27$ |
| Reflections collected | 57470 |
| Independent reflections | 2898 [$R_{int}$ = 0.0686, $R_{sigma}$ = 0.0229] |
| Data/restraints/parameters | 2898/0/85 |
| Goodness-of-fit on $F^2$ | 1.093 |
| Final R indexes [I>=2σ (I)] | $R_1$ = 0.0222, $wR_2$ = 0.0412 |
| Final R indexes [all data] | $R_1$ = 0.0304, $wR_2$ = 0.0437 |
| Largest diff. peak/hole / e Å$^{-3}$ | 3.19/-4.83 |

**Table S2 Extended.** Fractional Atomic Coordinates (×10$^4$) and Equivalent Isotropic Displacement Parameters (Å$^2$×10$^3$) of Batch 916 (Non-Tailored) $U_{eq}$ is defined as 1/3 of the trace of the orthogonalised $U_{IJ}$ tensor.

| **Atom** | ***x*** | ***y*** | ***z*** | **U(eq)** |
|---|---|---|---|---|
| Ir1 | 4113.8(2) | 0 | 3237.1(2) | 3.10(4) |
| Ir2 | 10287.7(2) | 0 | 1782.1(2) | 3.11(4) |
| Ir3 | 10000 | 0 | 0 | 2.91(5) |
| Ir4 | 5000 | 0 | 5000 | 3.18(5) |
| Ba5 | 7181.9(3) | 0 | 2515.9(2) | 4.59(5) |

**Table S2 Extended.** Fractional Atomic Coordinates (×10[4]) and Equivalent Isotropic Displacement Parameters (Å$^2$×10$^3$) of Batch 916 (Non-Tailored) $U_{eq}$ is defined as 1/3 of the trace of the orthogonalised $U_{IJ}$ tensor.

| Atom | *x* | *y* | *z* | U(eq) |
|---|---|---|---|---|
| Ba6 | 3488.0(3) | 0 | 779.2(2) | 4.40(5) |
| Ba7 | 1307.2(3) | 0 | 4280.9(2) | 4.68(5) |
| O1 | 10945(3) | 2289(5) | 970.5(19) | 5.3(4) |
| O2 | 8584(4) | 0 | 751(3) | 5.0(6) |
| O3 | 3820(3) | -2352(5) | 4168.2(19) | 5.7(4) |
| O4 | 6064(4) | 0 | 4009(3) | 5.9(6) |
| O5 | 12034(4) | 0 | 2744(3) | 6.2(6) |
| O6 | 9411(3) | -2546(5) | 2368(2) | 7.1(5) |

**Table S2 Extended.** Anisotropic Displacement Parameters (Å$^2$×10$^3$) for Batch 916 (Non-Tailored). The Anisotropic displacement factor exponent takes the form: $-2\pi^2[h^2a^{*2}U_{11}+2hka^*b^*U_{12}+\ldots]$.

| Atom | $U_{11}$ | $U_{22}$ | $U_{33}$ | $U_{23}$ | $U_{13}$ | $U_{12}$ |
|---|---|---|---|---|---|---|
| Ir1 | 3.01(7) | 3.00(7) | 3.44(7) | 0 | 1.08(5) | 0 |
| Ir2 | 2.74(7) | 3.17(7) | 3.45(7) | 0 | 0.83(5) | 0 |
| Ir3 | 2.66(9) | 3.19(10) | 2.97(10) | 0 | 0.86(7) | 0 |
| Ir4 | 2.76(10) | 3.01(10) | 3.88(10) | 0 | 1.00(7) | 0 |
| Ba5 | 3.96(10) | 4.01(11) | 6.08(11) | 0 | 1.74(8) | 0 |
| Ba6 | 3.85(11) | 3.79(11) | 5.25(11) | 0 | 0.45(8) | 0 |
| Ba7 | 5.03(11) | 3.89(11) | 5.83(11) | 0 | 2.73(9) | 0 |
| O1 | 8.4(11) | 3.9(11) | 4.1(10) | -1.2(8) | 2.3(8) | -2.4(8) |
| O2 | 4.7(15) | 8.1(16) | 2.0(14) | 0 | 0.4(11) | 0 |
| O3 | 7.2(11) | 6.0(11) | 3.7(10) | 0.6(9) | 0.7(8) | -1.7(8) |
| O4 | 5.5(15) | 7.9(16) | 4.9(15) | 0 | 2.5(11) | 0 |
| O5 | 3.8(14) | 8.9(17) | 5.2(15) | 0 | -0.5(11) | 0 |
| O6 | 8.5(11) | 4.8(11) | 8.5(12) | 3.1(9) | 2.8(9) | -0.5(9) |
| Ir1 | 3.01(7) | 3.00(7) | 3.44(7) | 0 | 1.08(5) | 0 |

**Table S2 Extended.** Bond Lengths for Batch 916 (Non-Tailored)

| Atom | Atom | Length/Å | Atom | Atom | Length/Å |
|---|---|---|---|---|---|
| Ir1 | Ir4 | 2.6234(3) | Ir4 | O3 | 2.029(3) |
| Ir1 | Ba5[1] | 3.4867(3) | Ir4 | O3[15] | 2.029(3) |
| Ir1 | Ba5 | 3.4856(4) | Ir4 | O3[16] | 2.029(3) |
| Ir1 | Ba5[2] | 3.4867(3) | Ir4 | O4 | 2.036(4) |

**Table S2 Extended.** Bond Lengths for Batch 916 (Non-Tailored)

| | | | | | |
|---|---|---|---|---|---|
| Ir1 | Ba6 | 3.6472(5) | Ir4 | O4[16] | 2.036(4) |
| Ir1 | Ba7 | 3.5274(4) | Ba5 | O1[1] | 2.851(3) |
| Ir1 | O3 | 2.026(3) | Ba5 | O1[17] | 2.851(3) |
| Ir1 | O3[3] | 2.026(3) | Ba5 | O2 | 3.305(4) |
| Ir1 | O4 | 2.028(4) | Ba5 | O3[18] | 3.065(3) |
| Ir1 | O5[4] | 2.038(4) | Ba5 | O3[7] | 3.065(3) |
| Ir1 | O6[2] | 2.000(3) | Ba5 | O4 | 2.755(4) |
| Ir1 | O6[5] | 2.000(3) | Ba5 | O5[1] | 2.8946(7) |
| Ir2 | Ir3 | 2.6605(3) | Ba5 | O5[2] | 2.8946(7) |
| Ir2 | Ba5[6] | 3.4724(3) | Ba5 | O6 | 2.714(3) |
| Ir2 | Ba5 | 3.5315(4) | Ba5 | O6[3] | 2.714(3) |
| Ir2 | Ba5[7] | 3.4724(3) | Ba5 | O6[5] | 3.065(3) |
| Ir2 | Ba6[7] | 3.5356(3) | Ba5 | O6[2] | 3.065(3) |
| Ir2 | Ba6[6] | 3.5356(3) | Ba6 | O1[1] | 2.861(3) |
| Ir2 | O1 | 2.013(3) | Ba6 | O1[19] | 3.244(3) |
| Ir2 | O1[3] | 2.013(3) | Ba6 | O1[4] | 2.932(3) |
| Ir2 | O2 | 2.030(4) | Ba6 | O1[9] | 3.244(3) |
| Ir2 | O5 | 2.000(4) | Ba6 | O1[17] | 2.861(3) |
| Ir2 | O6 | 2.013(3) | Ba6 | O1[20] | 2.932(3) |
| Ir2 | O6[3] | 2.013(3) | Ba6 | O2[1] | 2.8680(3) |
| Ir3 | Ba6[8] | 3.4092(4) | Ba6 | O2[10] | 2.733(4) |
| Ir3 | Ba6[9] | 3.5677(3) | Ba6 | O2[2] | 2.8680(3) |
| Ir3 | Ba6[6] | 3.5677(3) | Ba6 | O6[2] | 2.764(3) |
| Ir3 | Ba6[10] | 3.4092(4) | Ba6 | O6[5] | 2.764(3) |
| Ir3 | O1[11] | 2.036(3) | Ba7 | O3[5] | 2.882(3) |
| Ir3 | O1[12] | 2.036(3) | Ba7 | O3[13] | 2.834(3) |
| Ir3 | O1[3] | 2.036(3) | Ba7 | O3[21] | 2.834(3) |
| Ir3 | O1 | 2.036(3) | Ba7 | O3[2] | 2.882(3) |
| Ir3 | O2 | 2.014(4) | Ba7 | O3 | 2.888(3) |
| Ir3 | O2[11] | 2.014(4) | Ba7 | O3[3] | 2.888(3) |
| Ir4 | Ba7[13] | 3.4317(3) | Ba7 | O4[16] | 3.240(4) |
| Ir4 | Ba7[14] | 3.4317(3) | Ba7 | O4[1] | 2.8975(7) |
| Ir4 | Ba7[6] | 3.4317(3) | Ba7 | O4[2] | 2.8975(7) |
| Ir4 | Ba7[7] | 3.4317(3) | Ba7 | O5[4] | 2.605(4) |
| Ir4 | O3[3] | 2.029(3) | | | |

## IV. Supplementary Data on Reproducibility of Magnetic Behavior Across Batches

To further support the intrinsic nature of the field-induced transformations observed in $BaIrO_3$, we present additional *a*- and *c*-axis magnetic susceptibility $\chi_a$ and $\chi_c$ data collected from independently grown crystals corresponding to the M1, M2, and M2* synthesis conditions. These measurements were performed on multiple batches synthesized under identical magnetic field protocols. The results confirm the high reproducibility of the magnetic behavior reported in the main text, including the systematic suppression of the Néel temperature $T_N$ with increasing synthesis field strength. The close agreement across samples rules out sample-specific anomalies or stochastic impurity effects, reinforcing the conclusion that the observed phenomena are robust consequences of magneto-synthesis.

### A. Grown Crystals at the M1 Synthesis Conditions

**Supplementary Fig. 6:**

The *a*-axis magnetic susceptibility $\chi_a$ for two different samples from **Batch #353 (M1)**:

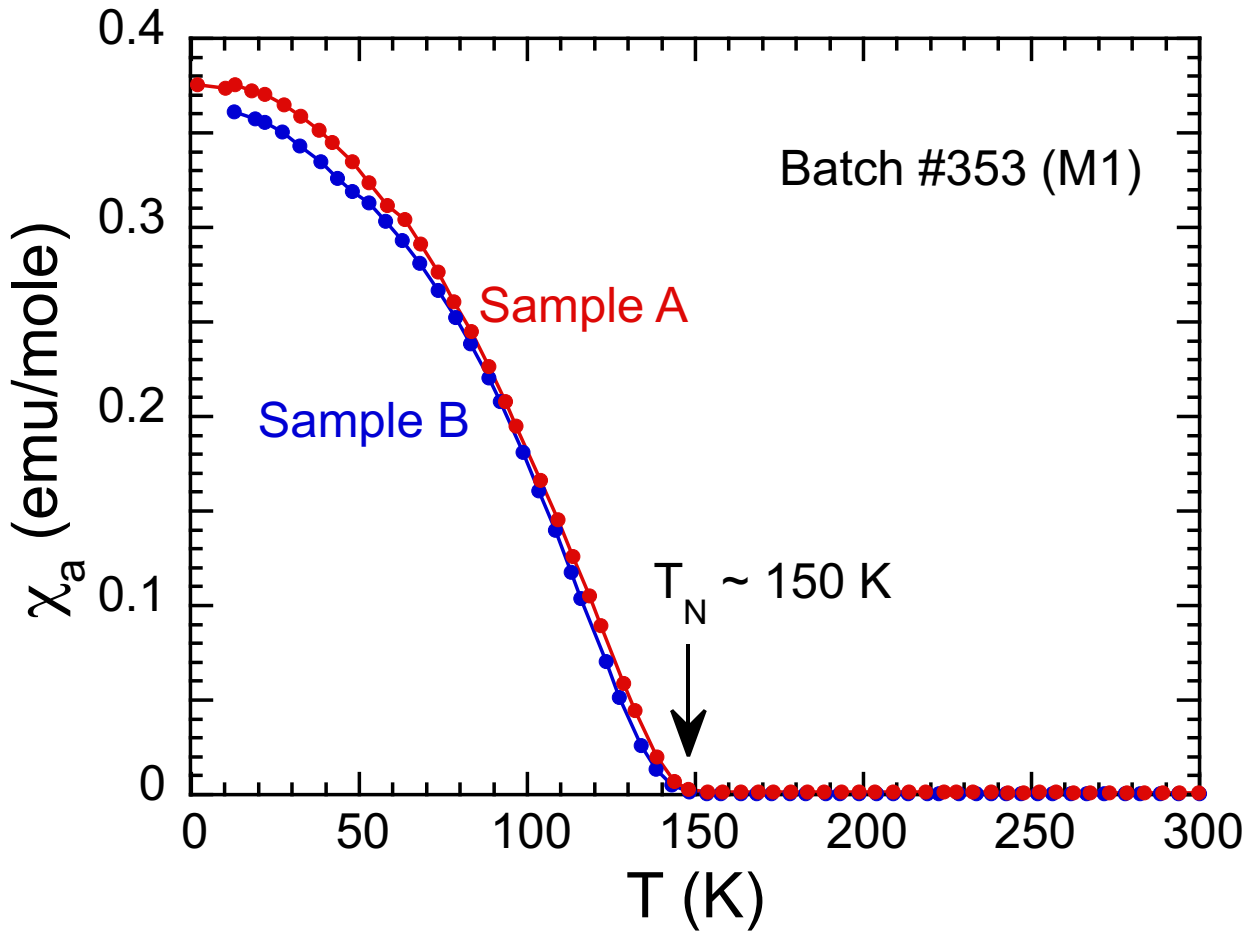

**Supplementary Fig. 7:**

The *c*-axis magnetic susceptibility $\chi_c$ for two different samples from **Batches #361 and 424 (M1)**:

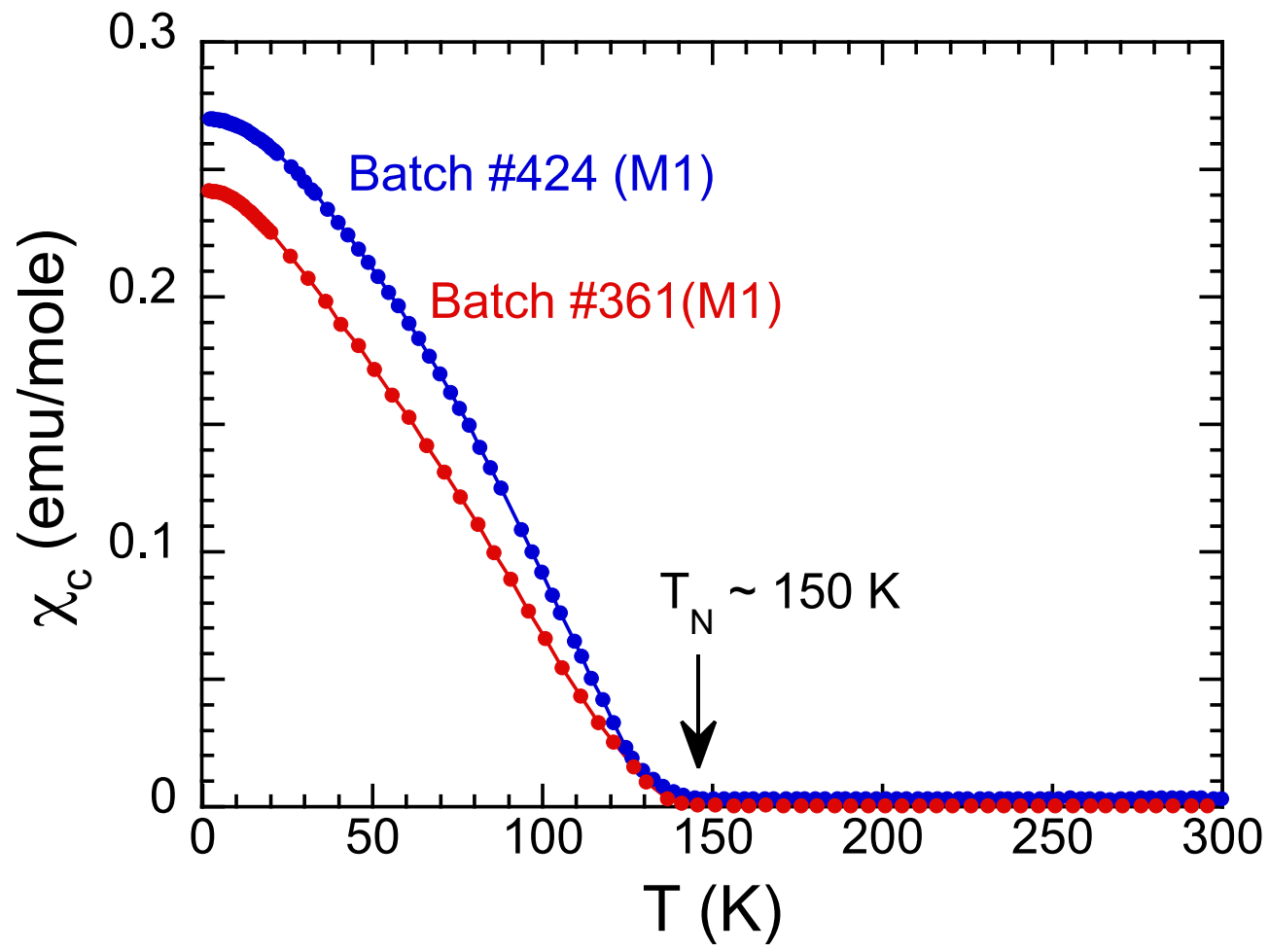

**B. Grown Crystals at the M2 Synthesis Conditions**

**Supplementary Fig. 8:**

The *a*-axis magnetic susceptibility $\chi_a$ for two different samples from **Batch #480 (M2)**:

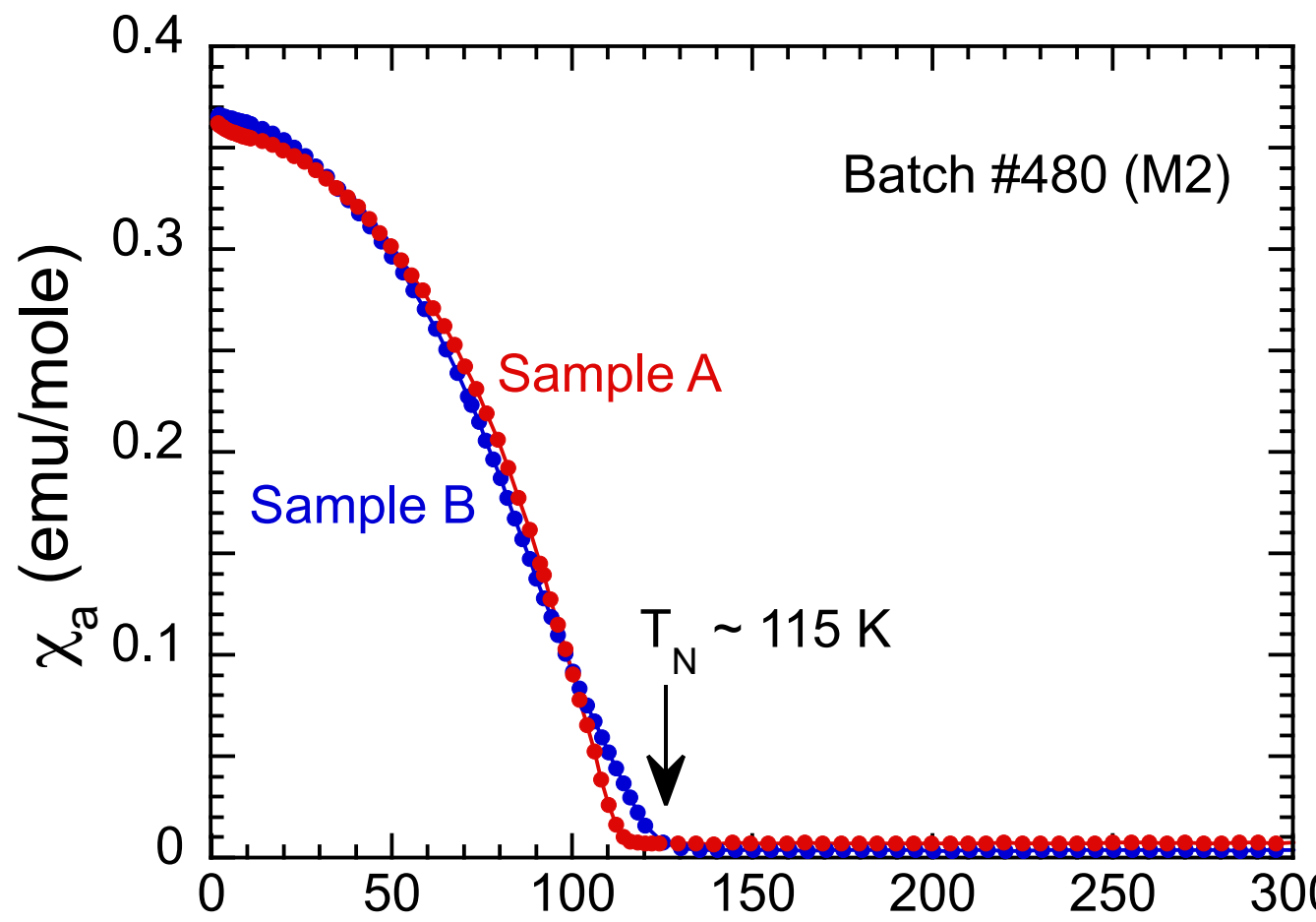

**Supplementary Fig. 9:**

The *c*-axis magnetic susceptibility $\chi_a$ for a sample from **Batch #469 (M2)**:

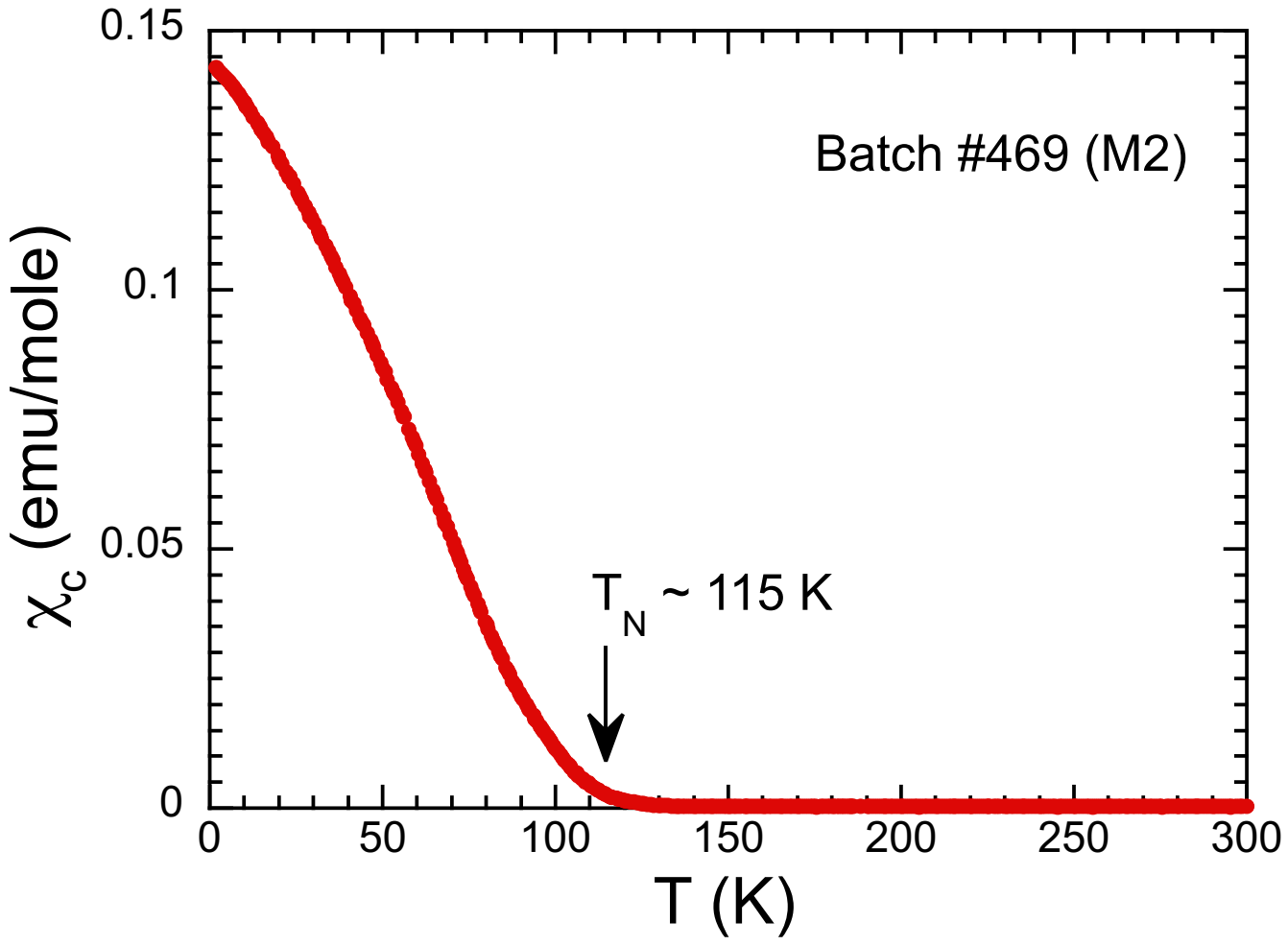

**C. Grown Crystals at the M2* Synthesis Conditions**

**Supplementary Fig. 10:**

The *c*-axis magnetic susceptibility $\chi_a$ for a sample from **Batch #470 (M2*)**:

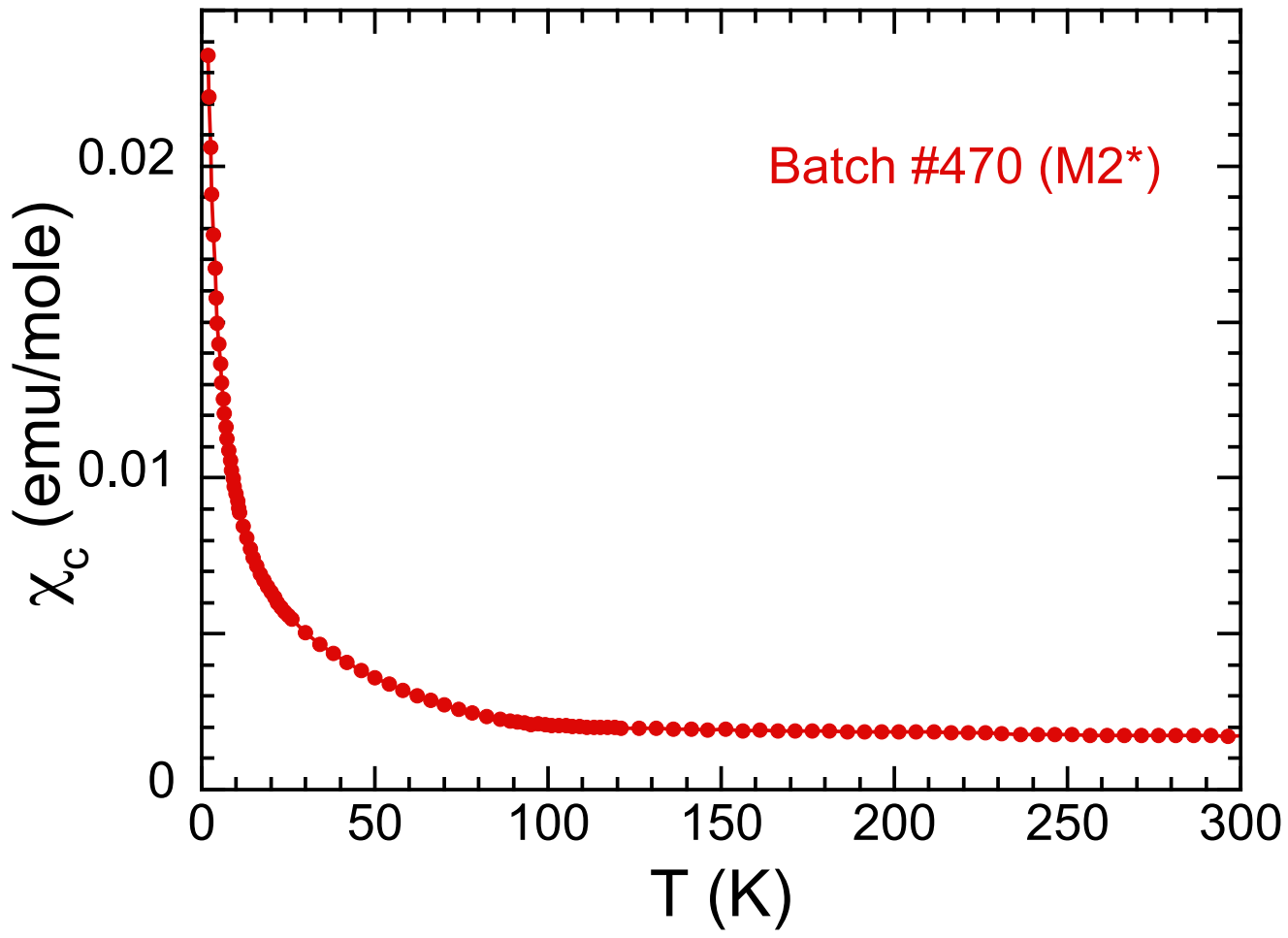